\documentclass[11pt,a4paper]{article}
\usepackage[margin=23mm,headheight=14pt]{geometry}
\usepackage[T1]{fontenc}
\usepackage[utf8]{inputenc}
\usepackage{amsmath,amssymb}
\usepackage{newtxtext,newtxmath}
\usepackage{microtype}
\usepackage{graphicx,booktabs,tabularx,array,xcolor}
\usepackage{longtable}
\usepackage{caption,float,placeins,listings}
\usepackage[numbers,sort&compress]{natbib}
\usepackage{xurl}
\usepackage[unicode,colorlinks=true,linkcolor=black,citecolor=black,
urlcolor=blue!45!black]{hyperref}
\hypersetup{
 pdftitle={Native Rectangular Linear Assignment with Implicit Rejection: A Jonker-Volgenant Solver without Square Expansion},
 pdfauthor={Ratha Siv, Raphael Reme, Matei Mancas, Bernard Gosselin},
 pdfsubject={Implementation and experimental study},
 pdfkeywords={linear assignment, Jonker-Volgenant, rectangular matching, implicit rejection}
}
\newfloat{algorithm}{tbp}{loa}
\floatname{algorithm}{Algorithm}
\title{Native Rectangular Linear Assignment with Implicit Rejection:\\
A Jonker--Volgenant Solver without Square Expansion}
\author{Ratha Siv\textsuperscript{1}, Raphael Reme\textsuperscript{2}, Matei Mancas\textsuperscript{1}, Bernard Gosselin\textsuperscript{1}\\[3pt]{\normalsize \textsuperscript{1}University of Mons, Belgium}\\{\normalsize \textsuperscript{2}Institut Pasteur, France}}
\date{15 November 2026}
\graphicspath{{latex/}{latex/images/}{images/}}
\makeatletter
\def\input@path{{latex/}}
\makeatother
\begin{document}
\maketitle

\begin{abstract}
Rectangular assignment with a uniform unmatched penalty is often implemented by expanding the cost matrix. We integrate established rectangular Jonker--Volgenant principles and implicit rejection with float64 working arithmetic and reusable workspaces in Rust (JVRUST), hybrid Rust/C++ (JVHYBRID) and C++ (JVCPP). A campaign evaluates laprust/lapcpp 1.2.0 and LAPX 0.13.0 in four Windows/Linux and Python environments. Each environment contains 302 cases: 3,286 complete-call timing conditions with five inputs and three rounds, plus 91 memory conditions with five fresh-process trials. Against the matching compact SciPy reference, all three native implementations exceed a 3\% advantage on all 102 required cases; JVRUST and JVCPP do so on 163--165 and 163--166 of 166 optional cases, respectively. Controlled comparisons qualify these gains by representation, initialization, sparsity, batching and orientation. We also assess how solver gains carry over to pedestrian and cell tracking using real-world data and prepared detections. In each Windows environment, the specialized LSAPE code wins 38 of 67 cases against the original Hungarian implementation and all 67 against its unchanged one-sided baseline. The contribution is software integration and reproducible performance characterization. Adverse results are retained; no algorithmic priority, statistical significance or universal solver ranking is claimed. The initialization policy was developed using related workload families, so the campaign is not a held-out test of heuristic optimality.
\end{abstract}

\FloatBarrier
\section{Introduction}

Multi-object trackers associate tracks with detections whose number can differ, while allowing an object on either side to remain unmatched. ByteTrack \cite{zhang2022bytetrack} and related association code express this through the \texttt{cost\_limit} argument of \texttt{lap.lapjv} \cite{gatagatlap}. Its padded implementation embeds the $N\times M$ costs in an $(N+M)$-square matrix. The limit prices rejection; it does not require maximizing the number of below-threshold pairs.

For float64 costs with 256 rows and 4096 columns, the original matrix occupies 8 MiB, square zero-padding requires 128 MiB, and optional-assignment expansion requires 144.5 MiB. These are cost-buffer dimensions, before workspace and output. Compact representations avoid this expansion, but their performance still depends on initialization, cost structure, orientation and interface work.

Rectangular JV, outsourcing and assignment with insertion and deletion costs already provide compact formulations and specialized searches \cite{jonker1987,volgenant1996,bijsterbosch2010,vanderwiel1997,crouse2016,bougleux2016,jones2017,bougleux2020}. Our contribution is to integrate these principles into Python-facing Rust, hybrid Rust/C++ and C++ implementations, then evaluate their complete calls against equivalent dense and sparse formulations. SciPy's rectangular solver already avoids required-assignment padding, and exact capped costs provide an original-dimension optional baseline. Neither dimensional savings nor implementation language alone explains the measured differences.

The paper contributes a self-contained derivation and optimality certificate for uniform-penalty assignment; implementations combining bounded reductions, a reusable rejection terminal and batch workspaces; and an empirical study with explicit representation, policy, work, orientation and application controls. Four environments retain separate measurements and a practical $\pm3\%$ tie band. The study includes unfavorable cases, planned coverage limits and the distinction between assignment calls, common-cost replay, pedestrian tracker updates and microscopy whole-linker latency. A separate LSAPE experiment examines related transformations within Hungarian code, without comparing JV and LSAPE runtimes.

Sections~\ref{sec:background}--\ref{sec:implementation} give the background, formulation and implementation. Section~\ref{sec:design} specifies the saved campaign, Section~\ref{sec:results} reports its results, and the discussion and limitations delimit their interpretation. Appendices organize additional evidence by workload and measurement question. The indexed supplement retains absolute values and provenance for every plotted observation.

\FloatBarrier
\section{Background and related work}\label{sec:background}

\subsection{Assignment algorithms and implementation lineage}

Jonker and Volgenant's algorithm combines initialization, reductions, and a specialized shortest augmenting path procedure \cite{jonker1987}. Initial stages often construct much of the matching cheaply. Remaining free rows require augmenting paths that can reassign existing pairs. A square matrix therefore does not eliminate the augmentation phase; it only removes the need for a dimensional transformation.

We use completion to mean the augmenting-path phase that finishes a partial matching. It does not refer to padding the cost matrix.

The baselines are established Jonker--Volgenant families, evaluated through their LAPX implementations \cite{lapx2026}. JVX (\texttt{lapjv}, \texttt{lapjvx}, \texttt{lapjvxa}) uses the gatagat/lap-derived dense core \cite{gatagatlap}; the dense campaign times \texttt{lapjvxa}, which returns assignment pairs. JVS (\texttt{lapjvs}, \texttt{lapjvsa}) uses the src-d/lapjv-derived implementation \cite{srcdlapjv}. These implementations have different initialization and numeric details. Their shared use of augmenting paths does not imply identical kernels.

SciPy \cite{virtanen2020scipy} supplies a native rectangular solver documented as a modified Jonker--Volgenant algorithm without initialization, referencing Crouse's treatment \cite{crouse2016,scipylsa}. Initialized rectangular JV also predates this work: Volgenant's RJV omits square column reduction and applies augmenting row reduction from zero column potentials \cite{volgenant1996}. Bijsterbosch and Volgenant refine RJV and specialize it for k-cardinality and outsourcing, including a capped-cost comparison on the original dimensions and a single relevant external option per person \cite{bijsterbosch2010}. That unexpanded outsourcing formulation originates with Vander Wiel and Sahinidis, whose shortest augmenting path algorithm treats external assignment through a single unrestricted arc \cite{vanderwiel1997}. These initialized rectangular and outsourcing searches are direct predecessors of the state and rejection handling studied here.

\subsection{Optional assignment and compact representations}

Optional assignment also has clear precedents. Bougleux and Brun formulate linear sum assignment with edition (LSAPE) using element-specific insertion and removal costs and a null element on each side \cite{bougleux2016}. Their adapted Hungarian algorithm permits augmenting paths to terminate at a null element, representing removal or insertion. The associated LSAPE toolbox supplies C++ implementations, including null-terminal path routines \cite{lsape2017}. The 2016 publication is the conceptual comparison; the July-2017 LSAPE-v4 release is the implementation used in our separate controlled experiment. Equality with the author-hosted v4 archive was checked on 27 October 2026, but does not establish identity with the code behind the 2016 timings.

Jones, Chawdhary, and King adapt Volgenant--Jonker methods to graph-edit matching \cite{jones2017}. Their compact representation avoids explicitly storing the redundant zero and infinite regions of the expanded matrix, using only (n+1)(m+1)-1 entries. Bougleux and colleagues subsequently describe a compact reduction from assignment with error correction to a rectangular assignment problem without cost constraints of the earlier reductions \cite{bougleux2020}.

Table~\ref{tab:lsape-design} compares Bougleux and Brun's published LSAPE design with our native JV implementation. With a uniform penalty, the two endpoint charges reduce to a single rejection price per unmatched smaller-side row, up to an additive constant (Section~\ref{sec:one-sided}). Within JV code descended from the JVS lineage, we combine rectangular reductions, rejection state stored as one distance and one predecessor row, the threshold arguments of \texttt{lap.lapjv}, and native batching. These established ingredients define a software-integration contribution; we claim no new optional-assignment objective, null-terminal principle, algorithmic priority or asymptotic improvement over the compact predecessors. Table~\ref{tab:methods} defines the measured interfaces and LSAPE arms. The separate experiment in Section~\ref{sec:lsape-results} evaluates one-sided adaptation within LSAPE-v4 and makes no JV-versus-LSAPE runtime comparison.

\begin{table}[!htbp]
\centering
\caption{Design choices in Bougleux and Brun's 2016 LSAPE model~\cite{bougleux2016} and in the native JV solver. Both avoid square expansion. For input dimensions $n,m$, the JV solver orients the costs with $r=\min(n,m)$ and $c=\max(n,m)$.}
\label{tab:lsape-design}
\begingroup\small
\setlength{\tabcolsep}{5pt}
\renewcommand{\arraystretch}{1.12}
\begin{tabularx}{\linewidth}{@{}>{\raggedright\arraybackslash}p{.16\linewidth}>{\raggedright\arraybackslash}X>{\raggedright\arraybackslash}X@{}}
\toprule
Design choice & Bougleux and Brun (2016) & Native JV solver (this work) \\
\midrule
Penalty model & Each element has its own removal or insertion cost. & Every unmatched endpoint costs $L/2$, equivalent in exact arithmetic to $L$ per rejected smaller-side row. \\
\addlinespace
Cost representation & An $(n+1)\times(m+1)$ edit-cost matrix includes a null row and a null column. & An $r\times c$ matrix stores the real costs. One scalar specifies the rejection price. \\
\addlinespace
Path endpoints & A path can end at an unassigned real element or a null element. & A path ends at a free real column or a reusable rejection terminal tracked by one distance and one predecessor row. \\
\addlinespace
Search & An adapted Hungarian algorithm finds augmenting paths for the edit assignment. & Bounded JV row reductions precede a shortest augmenting path search for the remaining free rows. \\
\addlinespace
Evaluation & The study times compact and square-expanded formulations on synthetic edit-cost matrices. & The Python interfaces support single solves and batches that reuse worker workspaces. Complete-call benchmarks include compact and expanded baselines. \\
\bottomrule
\end{tabularx}
\endgroup
\end{table}

Application interfaces also accept non-assignment costs. MATLAB's \texttt{assignjv}, \texttt{assignmunkres}, and \texttt{assignauction} take a scalar cost of non-assignment charged per unmatched endpoint \cite{mathworks2018assign}. The Tracker Component Library includes a Jonker--Volgenant routine for missed detections that never builds the augmented matrix but keeps search state for one virtual row per target \cite{crouse2017tcl}.

The compact reduction in \cite{bougleux2020} is also an executable comparison point. For uniform endpoint penalties, its cost factorization reduces, up to a constant shift, to capping real costs at the combined rejection price. Section~\ref{sec:compact} gives a self-contained argument for this specialization, including finite signed costs, and Section~\ref{sec:representations} evaluates it with SciPy. This provides a baseline without square expansion; it does not reproduce every optimization in the authors' LSAPE software.

\subsection{Interfaces and transformed problems}

JVRUST, JVHYBRID and JVCPP provide native required rectangular assignment and finite-limit rejection. JVX uses square padding or augmentation, and JVS receives a square problem through an adapter. SciPy receives the required, capped, dummy-column, diagonal-dummy or square-expanded representation named in each comparison. The required reference is ordinary rectangular SciPy; the optional reference is capped SciPy throughout the campaign.

The optional sparse comparisons include SciPy's \texttt{min\_weight\_full\_bipartite\_matching} (CSGRAPH) \cite{scipycsgraph} and LAPX's \texttt{lapmod} (LAPMOD). Their measured adapters construct admissible sparse edges and rejection structure from the same dense input. The layout adaptations credit Pylapy \cite{pylapy2026}; the benchmark source retains its MIT notice and identifies the adapted revision. Thus their complete-call times include representation construction. Table~\ref{tab:methods} distinguishes these numerical contracts from the separate LSAPE arms.

\FloatBarrier
\section{Problem formulation}\label{sec:formulation}

\subsection{Required rectangular assignment}

Let C be an N \ensuremath{\times} M real cost matrix, with forbidden edges excluded from the admissible edge set E. Orient the matrix so that r = min(N,M) and c = max(N,M), with r \ensuremath{\leq} c. In (1) and (4), \(x_{ij}=0\) for \((i,j)\notin E\). A required rectangular assignment solves

\[
\min_x \sum_{(i,j)\in E} C_{ij}x_{ij},
\quad
\sum_j x_{ij}=1,\quad
\sum_i x_{ij}\leq 1,\quad
x_{ij}\in\{0,1\}.
\tag{1}
\]

The solver must match all r rows if this is feasible. Surplus columns remain unused. If the original matrix is tall, the returned indices are mapped back to the original orientation.

\noindent\begin{minipage}{\linewidth}
\subsection{Finite-limit assignment}\label{sec:finite-limit}

Let L be a finite limit and A a matching with K pairs. The finite-limit objective is

\[
F(A)=\sum_{(i,j)\in A}C_{ij}
     +\frac{L}{2}(r+c-2K).
\tag{2}
\]
\end{minipage}

Each unmatched vertex incurs L/2. In the square augmentation used by JVX, the real costs occupy the upper-left block. Real-to-dummy and dummy-to-real blocks cost L/2, while the dummy-to-dummy block is zero.

Equation (2) is not equivalent to solving a required assignment and then discarding expensive pairs. For example, let

\[
C=\begin{pmatrix}0&0.9\\0.9&+\infty\end{pmatrix},\qquad L=1.
\]

Two real pairs cost 1.8. Selecting only the zero-cost pair leaves two unmatched vertices and costs 1.0 under (2). Both nonzero finite edges are below the threshold, yet the smaller matching is preferable.

At equality, a real edge of cost L can tie with rejecting both endpoints. The objective does not impose maximum cardinality among equal-cost optima. Applications that require a particular inclusive-threshold or cardinality rule must specify it separately.

\subsection{One-sided rejection equivalence}\label{sec:one-sided}

\noindent\begin{minipage}{\linewidth}
\textbf{Proposition 1.} In exact arithmetic, minimizing (2) is equivalent to minimizing

\[
P(A)=\sum_{(i,j)\in A}C_{ij}+L(r-K)
\tag{3}
\]

\end{minipage}
on the oriented real rectangular problem.

\textbf{Proof.} Rearranging the rejection term gives

\[
F(A)=\sum_{(i,j)\in A}C_{ij}+L(r-K)+\frac{L}{2}(c-r).
\]

The final term is independent of A. Consequently, F and P have the same minimizers. The argument holds for any finite real L, including zero and negative values, and includes empty dimensions. \hfill\ensuremath{\blacksquare}

The transformation is useful because every smaller-side row can either select a real column or take a rejection option costing L. No separate dummy column is needed for each possible rejection. The constant must be restored when comparing the full penalized objective, but it does not affect the optimizer.

Signed support is an interface convenience rather than a new uniform-penalty model: adding a constant \(k\) to all admissible real costs and to \(L\) shifts \(P\) by the constant \(kr\) in exact arithmetic. Floating-point execution still requires the numeric contract below.

The native interfaces report the sum of the selected costs when requested. This total excludes rejection penalties and uses the input conversion described in Section~\ref{sec:numeric}. Validation therefore reconstructs the penalized objective from the returned pairs.

\subsection{Primal-dual certificate}

For an effective row penalty p, introduce a rejection variable \(z_{i}\):

\[
\begin{aligned}
\min_{x,z}\quad&
\sum_{(i,j)\in E}C_{ij}x_{ij}+p\sum_i z_i\\
\text{subject to}\quad&
\sum_j x_{ij}+z_i=1,\qquad
\sum_i x_{ij}\leq1,\qquad x_{ij},z_i\geq0.
\end{aligned}
\tag{4}
\]

This is a network-flow formulation: each row contributes one unit, real columns have unit capacity, and a row may send its unit directly to the terminal at cost p. The integral capacities yield integral optimal solutions. Its dual is

\[
\max_{u,v}\sum_i u_i+\sum_jv_j,
\quad
u_i+v_j\leq C_{ij}\quad((i,j)\in E),\quad u_i\leq p,\quad v_j\leq0.
\tag{5}
\]

\textbf{Proposition 2.} A feasible matching/rejection solution is optimal if a feasible dual satisfies tightness on matched real edges, \(u_{i}=p\) for rejected rows, and \(v_{j}=0\) for unused columns.

\textbf{Proof.} The primal-dual gap can be written as

\[
P-D=
\sum_{(i,j)\in E}(C_{ij}-u_i-v_j)x_{ij}
+\sum_i(p-u_i)z_i
+\sum_jv_j\left(\sum_i x_{ij}-1\right).
\tag{6}
\]

Every term is nonnegative by primal and dual feasibility. The stated conditions make each term zero, so the primal and dual objectives coincide. Weak duality establishes optimality. \hfill\ensuremath{\blacksquare}

For required rectangular assignment, rejection variables and the constraints \(u_{i}\leq p\) are omitted. The non-positive column potentials and zero potentials on unused columns remain important. Ordinary square reductions cannot be transferred blindly to a rectangular implementation.

These propositions establish the formulation and a final optimality certificate. They are not a formal verification of every implementation transition or a floating-point error bound.

\subsection{Compact equivalent formulations}\label{sec:compact}

The one-sided objective also permits two formulations for an ordinary required rectangular solver. First, append r dummy columns of cost L to the oriented real matrix. The resulting r \ensuremath{\times} (c+r) problem assigns every row either to a real column or to one of the distinct dummy columns. Removing dummy pairs recovers an optimal finite-limit matching under (3).

Second, no additional columns are necessary when rejection costs are uniform. Define a capped matrix on the original r \ensuremath{\times} c dimensions:

\[
\widehat C_{ij}=\min(C_{ij},L),
\qquad \min(+\infty,L)=L.
\tag{7}
\]

\textbf{Proposition 3.} Let B be a minimum-cost required matching of all r rows under (7). Retain only its pairs whose original cost is strictly below L. The retained matching minimizes (3); rejecting equality ties is one valid optimality convention.

\textbf{Proof.} For any required matching B, let A be its retained subset. Every discarded pair costs L in the capped problem, so the capped cost of B is exactly P(A). Conversely, any finite-limit matching A can be extended to r pairs because c\ensuremath{\geq}r. Each of the added pairs costs at most L after capping, and each original pair costs at most its original cost. Thus some required matching has capped cost at most P(A). Taking minima in both directions proves equality of the optimal values and optimality of the retained subset. The argument holds for finite real L and finite admissible costs, including negative values; forbidden original pairs are discarded. \hfill\ensuremath{\blacksquare}

This is the uniform-penalty case of the substitution factorization underlying \cite{bougleux2020}, stated here to make the comparison reproducible. Under this specialization, Bougleux et al.\ (Theorem 1, Eqs. (7)--(9)) retain selected substitutions at equality, whereas Proposition 3 and the capped SciPy adapter discard them. Both conventions preserve the optimal penalized objective in exact arithmetic, but may return different pairs and cardinalities. Capping occurs before optimization. It is therefore different from solving the original required problem and then discarding expensive pairs, which fails in the example in Section~\ref{sec:finite-limit}. The capped and dummy-column baselines are equivalent in exact arithmetic; their floating-point behavior is assessed separately by the validation experiments.

Table~\ref{tab:methods} gives the representation dimensions and objectives. A reusable terminal is a search option, not a capacity-one dummy column. These dimensions describe buffers, not measured process memory.

\FloatBarrier
\section{Implementation}\label{sec:implementation}

\subsection{Shared architecture and three implementations}

Each implementation consists of a thin Python extension and a native core. The extension validates shape, real numeric type, precision, and limit options, prepares a contiguous working representation where necessary, and invokes the core. The single-matrix and batch interfaces, return contracts, and error behavior are identical across the three implementations. JVRUST is provided by laprust 1.2.0 and the \texttt{lapjvr} family in LAPX 0.13.0; JVHYBRID by LAPX \texttt{lapjvh}; and JVCPP by lapcpp 1.2.0 \cite{laprust2026,lapx2026,lapcpp2026}.

Interface compatibility here means that the single-matrix entry points accept the \texttt{extend\_cost}, \texttt{cost\_limit}, and \texttt{return\_cost} arguments of \texttt{lap.lapjv} with the same objective: the finite limit prices unmatched endpoints, and the reported total excludes those penalties. They return matched row and column index arrays, or pairs, rather than lap's full-length per-row and per-column assignment vectors, so callers of lap need a small output adaptation. Compatibility does not imply identical assignments when optima tie, or identical behavior across precision modes. At \(C_{ij}=L\), matching and rejection tie in exact arithmetic. Native reductions and search prefer rejection on computed ties, whereas the evaluated padded \texttt{lap.lapjv} can retain the edge. This is not a strict \(C_{ij}<L\) output guarantee: the rounded effective penalty \(p\) in Section~\ref{sec:numeric} can differ from the supplied \(L\).

The native dispatcher selects the square kernel only for square inputs with an infinite limit. The square kernel's initialization and reductions are followed by an augmentation phase that completes the matching. JVRUST executes this phase in Rust. JVHYBRID calls a C++ completion kernel that operates on Rust-owned workspace and preserves the completion strategy and operation order. JVCPP implements the whole algorithm in C++, including the same C++ completion routine. The campaign times JVRUST through \texttt{laprust.lapjv} and JVHYBRID through LAPX \texttt{lap.lapjvh}. Their required-square calls use different completion kernels, but package and build also change. JVRUST and JVCPP compare complete Rust and C++ implementations of the same algorithm.

All rectangular inputs and finite-limit inputs, including square matrices with a finite limit, use the rectangular core. JVRUST and JVHYBRID share that Rust core, so differences between them on these paths cannot be interpreted as evidence of different rectangular algorithms. JVCPP executes the same rectangular algorithm in C++.

\subsection{Native rectangular reductions and search}

The rectangular core stores separate mappings, potentials, predecessors, distances, and free-row lists. Following Volgenant's RJV initialization \cite{volgenant1996} and its later refinements \cite{bijsterbosch2010}, it omits square column reduction and starts column potentials at zero. Its bounded augmenting row reductions lower a column potential only while assigning that column, maintaining non-positive potentials and zero potentials on unused columns.

After reductions, each remaining free root is introduced with a feasible row potential obtained from its available reduced costs and, in finite-limit mode, the rejection penalty. The root's reduced costs are retained as the initial shortest-path distances, so the first relaxation does not read the cost row again. A shortest-path search then scans real columns. Reaching a free column completes an augmenting path. Reaching an occupied column continues through its assigned row, allowing previous pairs to be rearranged.

The implementations use two bounded rectangular row-reduction passes with \(B=\min(c,4)\). Each pass permits at most B times its initial free-row count in row scans, including displacement retries. Among equally cheap real columns, a free column is preferred \cite{bijsterbosch2010}; in finite-limit mode the implicit rejection option still wins its computed ties. For required rectangular matching, \(\max(r,8)\) consecutive displacements without acquiring a free column disable further price lowering. Acquiring a free column resets the streak until the cutoff is reached; once disabled, the cutoff persists across both passes. The current scan completes normally. Finite-limit inputs, including optional squares, use the same scan allowance and tie preference without that stall cutoff. Remaining free rows enter augmenting search. These rules limit initialization work without removing admissible edges or changing the objective. Required square problems retain their separate square initializer, which samples four complete rows to select an existing initialization strategy.

Policy selection used development workload families, including congested and tied costs, and is not a held-out validation of an optimal heuristic. The required-only cutoff limits unproductive displacement chains while allowing optional chains to continue within the scan allowance. Free-column tie scans can add work on heavily tied costs. Separately built policy variants test these choices within matched implementations (Section~\ref{sec:policy}); they do not assign a causal share of the native-versus-SciPy difference to any rule.

\subsection{Implicit rejection terminal}

For finite-limit matching, the search compares real-column paths with a rejection terminal. A real edge has reduced cost \(C_{ij}-u_i-v_j\). Leaving a row unmatched has reduced cost \(p-u_i\). The implementation stores the best terminal distance and its predecessor row using the variables \texttt{dummy\_distance} and \texttt{dummy\_row}.

More precisely, let \(\delta_i\) be the shortest-path distance at which row i is reached; the root has \(\delta=0\). When row i is reached, the rejection candidate \(\delta_i+p-u_i\) updates the terminal distance \(D_\perp\), the smallest such value so far, and its row. Each unsettled real column j is relaxed to \(\min(d_j,\ \delta_i+C_{ij}-u_i-v_j)\). If \(D_\perp\) is no greater than the smallest unsettled column distance, rejection is selected; otherwise that column is settled, and the search either ends at a free column or continues through the column's matched row. The terminal therefore adds one distance and one predecessor row to the ordinary search state.

Let D be the distance of the selected endpoint. The root's potential increases by D. Every settled column j has its potential lowered by \(D-d_j\) and, if matched, its row's potential raised by the same amount. Assignments are then reversed from the free column, or from the terminal's predecessor row, back to the root. A free column ending the path has \(d_j=D\) and keeps its zero potential, so column potentials stay non-positive and unused columns stay at zero.

If rejection is selected, the number of real pairs need not increase: the path can replace one row's real assignment and leave another row unmatched.

\begin{figure}[!htbp]
\centering
\includegraphics[width=\linewidth,height=.50\textheight,keepaspectratio]{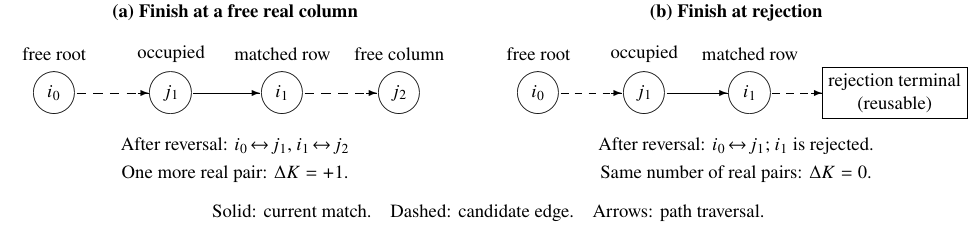}
\caption{Conceptual paths to a free real column or reusable rejection terminal. A free-column endpoint adds a real pair; a rejection endpoint can transfer unmatched status without increasing cardinality. This illustrates the algorithm rather than an observed execution trace.}
\label{fig:paths}
\end{figure}

Figure~\ref{fig:paths} illustrates the two endpoints, and Algorithm~\ref{alg:1} places them within the complete rectangular solve.

\begin{algorithm}[!htbp]
\caption{High-level native rectangular control flow with implicit rejection. The implementation additionally specifies reduction budgets, finite-value checks, tie handling, and workspace reuse.}
\label{alg:1}
\begin{lstlisting}
Orient C so that r <= c.
Initialize mappings, potentials, and free-row state.
Apply bounded row reductions, including rejection choices in finite-limit mode.
For every row still requiring processing:
    Establish the root row potential from real reduced costs and finite p.
    Search shortest alternating paths over real columns.
    In finite-limit mode, also track the best rejection terminal.
    Stop at a free real column or the selected rejection terminal.
    Update potentials and reverse assignments along the path.
Map real pairs back to the original matrix orientation.
\end{lstlisting}
\end{algorithm}

\FloatBarrier
\subsection{Storage, batching, and operation bounds}

For contiguous wide inputs in the selected cost dtype, the real cost matrix can be borrowed. Solver state is \(\mathcal{O}(r+c)\), with float64 numerical workspace even when costs are float32; widening this linear workspace does not require a dense float64 cost copy. Pair output is \(\mathcal{O}(r)\). Tall inputs may require an \(\mathcal{O}(rc)\) orientation copy, and dtype or layout conversion can also require a dense copy. Including these costs, storage remains \(\mathcal{O}(rc+r+c)\), rather than requiring a c-square or (r+c)-square cost matrix.

Batch input storage is shared with the call. Each active worker retains its own workspace and can reuse it across problems; results are collected in input order, and the lowest failing index is reported on error. JVRUST and JVHYBRID schedule problems on reusable Rayon thread pools. JVCPP uses a reusable C++ thread pool and claims consecutive small problems in chunks, bounded to about 4096 costs per claim so that adjacent difficult problems are not serialized on one worker. Wide inputs require no per-worker padded cost matrix; tall problems can require orientation buffers. This distinction is relevant to peak memory as worker counts increase.

Each bounded row-reduction pass performs \(\mathcal{O}(rB)\) scans of at most c real costs; the additional free-column tie scan has the same order. Since \(B\leq4\), the two passes cost \(\mathcal{O}(rc)\). Each final augmenting search visits at most r occupied real columns before reaching a terminal, scanning up to c candidates at each step. Across r roots, augmentation costs \(\mathcal{O}(r^2c)\). Thus \(\mathcal{O}(rc+r^2c)\) is a conservative bound for this release's rectangular solve, excluding separately described conversion and output work. The required-only cutoff can reduce initialization work but is unnecessary for this bound. This is an implementation bound, not an asymptotic improvement over established rectangular assignment methods.

\subsection{Numeric and interface contract}\label{sec:numeric}

All three native implementations retain selected float32/float64 cost storage while using float64 potentials, distances, minima, and reduced-cost arithmetic on required square, rectangular, and finite-limit paths. LAPX's JVS implementation also widens its numerical working state. The wider working arithmetic reduces cancellation when large costs coexist with small assignment differences; it cannot restore distinctions lost during input conversion or guarantee exact optimization for arbitrary dynamic ranges. Finite float64-to-float32 conversion overflow is checked. Other accepted real numeric dtypes first undergo NumPy conversion to float64 without an equivalent explicit overflow check; on platforms with wider real types, a finite value outside binary64 range can become positive infinity at that stage. A motivating upstream report \cite{lapjvprecision2026} gives the exactly representable float32 matrix \(\left(\begin{smallmatrix}2^{24}&2^{24}\\1/8&1/4\end{smallmatrix}\right)\): choosing the diagonal costs \(2^{24}+1/4\), while the off-diagonal costs \(2^{24}+1/8\). Accumulating selected original costs in float64 exposes the difference that float32 arithmetic can conceal. This witness motivates regression coverage without establishing robustness for every numerical scale.

For a finite limit \(L\), the interface rounds \(L/2\) to the selected cost dtype \(T\), doubles that value in \(T\), and converts the result to the float64 row penalty \(p\). The effective row penalty p need not equal the supplied L. The theoretical equivalence therefore applies to the represented costs and effective penalty. The controlled performance study uses float64 inputs; the pedestrian study retains float32 tracker costs and separately widens the same values for replay (Appendix~\ref{app:applications}). Saved validation checks the represented objective and targeted numerical and interface cases; it does not establish arbitrary-range exactness. Native output does not apply a raw \(C_{ij}<L\) post-filter, so computed-tie preference does not guarantee rejection whenever the raw cost equals the supplied limit. When requested, the native interfaces sum selected costs in float64 from the input buffer before any precision change requested for the solve. That buffer retains float32 or float64 inputs and converts other accepted real dtypes to float64 first.

Positive infinity represents a forbidden edge. NaN and negative infinity in the matrix are outside the documented supported contract. Required matching can be infeasible, and reduced-cost overflow can also cause an error; these conditions should not be conflated. As defensive checks, a path reversal that does not reach its root and a final matching whose row and column mappings disagree are reported as internal errors rather than returned.

All methods in the controlled assignment timing suites return int64 pair arrays and omit total-cost calculation. Pair ordering remains native to each interface. Untimed validation checks feasibility and objective values, rather than demanding identical arrays when optimal assignments tie.

\FloatBarrier
\section{Experimental design}\label{sec:design}

\subsection{Environment, placement and measurement boundary}

The light campaign contains one session in each environment: Linux/Python 3.11.18 (L311), Linux/Python 3.14.10 (L314), Windows/Python 3.11.18 (W311), and Windows/Python 3.14.10 (W314). This order is used throughout. All four October sessions measure laprust 1.2.0, lapcpp 1.2.0, LAPX 0.13.0, NumPy 2.4.6 and SciPy 1.17.1. The separate November microscopy session uses LAPX 0.13.1; its measurements are not pooled with these environments. The shared dependency versions avoid confounding the interpreter comparison with different NumPy/SciPy releases. As of 31 October 2026, SciPy 1.17.1 is the newest non-yanked stable release whose declared Python requirement admits both interpreters; 1.18.1 requires Python $\geq3.12$ \cite{scipyversions2026}. The rectangular assignment source/header and sparse matching source are byte-identical between those two tags \cite{scipysource2026}; this source check does not establish identical builds or timings. Version numbers and native-module hashes are recorded per environment; the comparisons do not isolate interpreter or operating-system causality.

Both operating-system records identify an Intel Core i9-13900KS. The launcher verifies that logical CPUs 0, 2, 4 and 6 belong to four distinct performance cores; single-threaded work uses CPU 0. Windows records the Balanced power scheme and Linux the powersave governor. Clock frequency and background activity are observed rather than fixed. The saved topology establishes physical-core separation for the selected logical CPUs.

One command runs assignment, diagnostics and application stages, with Windows additionally running the frozen LSAPE study. Recorded total launcher times are 66.23, 65.39, 85.71 and 85.32 minutes in L311/L314/W311/W314, including the recorded preparation state and suite stages. These are completed elapsed times on the recorded configurations, including their existing preparation state; they do not predict a clean-machine build time. Primary timings cover complete calls: input conversion and representation construction where required, allocation/workspace handling, solving and pair output. For optional assignment, timing includes a common output filter that also applies to native results. It retains only pairs whose original costs are finite, satisfy any gate and are strictly below $L$. Input generation and untimed reference validation are outside the timing boundary. Constructor-only and instrumented measurements are secondary evidence and are never substituted for complete-call latency.

\begin{table}[!htbp]
\centering
\caption{Protocol and evidence inventory for the four October environments. Repeated rounds are within-session observations; timing and memory conditions have different contracts. The November microscopy session is separate.}
\label{tab:protocol}
\begingroup\small
\setlength{\tabcolsep}{3pt}
\renewcommand{\arraystretch}{1.12}
\begin{tabular}{lcccc}
\toprule
Attribute & L311 & L314 & W311 & W314 \\
\midrule
Python & 3.11.18 & 3.14.10 & 3.11.18 & 3.14.10 \\
Operating system & Linux & Linux & Windows & Windows \\
Single/batch logical CPUs & \multicolumn{4}{c}{0 / 0, 2, 4, 6; distinct physical performance cores} \\
laprust / lapcpp / LAPX & \multicolumn{4}{c}{1.2.0 / 1.2.0 / 0.13.0} \\
NumPy / SciPy & \multicolumn{4}{c}{2.4.6 / 1.17.1} \\
Inputs / rounds / target & \multicolumn{4}{c}{5 / 3 / 20 ms per dense timing block} \\
Single required / optional & \multicolumn{4}{c}{102 / 166 cases per environment} \\
Batch/workspace / memory & \multicolumn{4}{c}{14 / 20 cases per environment} \\
Completed/planned timing & 3,286/3,286 & 3,286/3,286 & 3,286/3,286 & 3,286/3,286 \\
Memory conditions / trials & \multicolumn{4}{c}{91 / 5 fresh processes per condition} \\
Application pipelines / rounds & \multicolumn{4}{c}{8 / 3; 999 frames per pipeline} \\
Captured matrices / precisions & \multicolumn{4}{c}{7,539 / float32 and widened float64} \\
Replay block target / warmup & \multicolumn{4}{c}{30 ms / 32 tracker frames} \\
LSAPE cases / processes & --- & --- & 67 / 3 & 67 / 3 \\
LSAPE inputs / rounds / target & --- & --- & 4 / 9 / 30 ms & 4 / 9 / 30 ms \\
\bottomrule
\end{tabular}

\endgroup
\end{table}

\subsection{Workload coverage and estimators}

Each environment has 302 cases: 268 single matrices (102 required and 166 optional), 14 batch/workspace workloads and 20 memory cases. Of the 3,377 planned conditions, 3,286 measure timing and 91 measure memory. All are complete, with no malformed records or unobserved planned timing conditions. Geometry contributes 120 single-matrix cases; the remaining families cover base inputs, formulations, size and aspect-ratio scaling, large and extreme shapes, congestion, forbidden entries, ties, signed costs and rejection limits.

Non-geometric synthetic costs start from independent $U_{ij}\sim\mathrm{Uniform}[0,1)$. Signed costs use $2U-1$. The congested-block family adds 3 to the two off-diagonal blocks formed by the first $3r/4$ rows and $c/4$ columns (integer floors); the congestion sweep varies the row fraction over $0.25,0.5,0.625,0.75,0.9$ at column fraction $0.25$. Quantized costs use $\lfloor kU\rfloor/k$ for $k\in\{1,2,8,256\}$. At $k=1$, all five inputs are the same all-zero matrix for both required and optional assignment. The forbidden-entry sweep masks independent entries with probabilities $0,0.5,0.9,0.99$ while retaining a feasible diagonal. Its optional limit is $0.00625$, compared with $0.5$ for congestion and $0.25$ for the tie sweep. Individual base and scaling cases retain their manifest-specific parameters.

The frozen generator preserves legacy PCG64 streams and offsets; it does not derive every input from one common seed. Extended families use seed 20261024 with a group-key hash and input index; geometry and supplemental inputs use seed 20261030 with the full SHA-256 group-key digest and input index. The manifests and source snapshots retain the legacy seeds, stream offsets and exact construction rules. Input hashes identify the actual measured corpus.

Every assignment timing condition uses five frozen inputs, three rounds and a 20 ms target block. The estimate is the median of three round means, with equal weight to each input within a round. Single-matrix ratios divide reference and method estimates from the same successful paired attempt; they are not medians of input-wise ratios. Batch comparisons within one placement also use the same attempt. Worker-count contrasts compare matched inputs in separate process attempts under their respective CPU placements. Required cases use ordinary SciPy and optional cases use capped SciPy as their primary comparator. JVHYBRID's canonical coverage is required-only; its absence in optional panels is planned.

Ratios below 0.97 are losses, values in $[0.97,1.03]$ are practical ties, and values above 1.03 are wins. This band is descriptive. We neither pool environments nor interpret timing-round spread as independent-session uncertainty. Derived tall/wide contrasts also join distinct matched-corpus case attempts. All estimates, references, input identities, attempts, status and reasons remain in the indexed supplement.

\subsection{Geometry, representations and controlled policies}

The tracking-like geometry design is inspired by Pylapy \cite{pylapy2026}. It draws $r$ source points uniformly in the unit square, selects $\mathrm{round}(qr)$ survivors without replacement, and perturbs each surviving coordinate by an independent zero-mean Gaussian with standard deviation $m\delta$, with motion multiplier $m\in\{0.25,1\}$. It adds $c-\mathrm{round}(qr)$ new uniform points, shuffles the observations, and uses Euclidean distances as costs. Moved points are not clipped or wrapped back into the square. Tracking-like geometry spans 8--4096 rows with square and $5:4$ shapes. The size sweeps retain two settings at every size: high survival/low motion ($q=0.9$, $m=0.25$) and low survival/high motion ($q=0.6$, $m=1$), where $\delta=r^{-1/2}$ is the generator's characteristic spacing. The complete $256\times320$ and $1024\times1280$ panels retain all four survival/motion settings. The former also varies rejection limit and gate, including the primary cases deduplicated into that grid. Connections in the figures hold the generator, objective, limit policy, orientation and environment fixed apart from the named sweep variable.

In geometry, JVX, JVS and square-expanded SciPy are not planned above 1024 rows. This also stops unpadded required-square JVX/JVS. It is an additional light-profile rule, distinct from the general admission limit on padded buffers of $4096^2$ entries. Missing points beyond a declared cutoff are not timing failures or evidence of a timeout. The native and matching compact SciPy comparisons continue to the largest sizes. Appendix~\ref{app:sensitivity} examines the protocol reduction using a separate saved full session, without adding its timings to the primary results.

The formulation and padded-layout panels separate mathematical representation from placement of real and dummy costs. The controlled policy panel uses separately built Rust and C++ variants with matched costs and output contracts. It varies bounded reduction budgets, bypasses reductions, disables free-column tie preference, and disables the required-only stall cutoff where applicable. Primary latency and separately instrumented work are joined by case, policy and input identity. Work-event counts are descriptive operations, not interchangeable units of processor time.

\begin{table}[!htbp]
\centering
\caption{Methods, representations and controlled LSAPE arms. Here $r\le c$ denotes oriented JV dimensions; LSAPE retains original $n,m$. Definitions are not measured memory or speed claims.}
\label{tab:methods}
\begingroup\small
\setlength{\tabcolsep}{3pt}
\renewcommand{\arraystretch}{1.12}
\begin{tabular}{>{\raggedright\arraybackslash}p{.20\linewidth}>{\raggedright\arraybackslash}p{.37\linewidth}>{\raggedright\arraybackslash}p{.34\linewidth}}
\toprule
Identity & Implementation / numerical contract & Representation or control \\
\midrule
JVRUST; JVCPP & Rust; C++; float64 work for float32/64 storage & Required or uniform rejection; original costs plus linear state. \\
JVHYBRID & Rust with C++ square completion; canonical required coverage & Shares JVRUST's optional path; not separately timed there. \\
JVX; JVS & LAPX pairs API; square JV & Required square padding when needed; optional square augmentation. \\
CSGRAPH; LAPMOD & SciPy sparse matcher; LAPX sparse solver & Optional sparse adapter includes graph construction and output mapping. \\
SciPy required & Modified rectangular JV; float64 & Original $r\times c$ required problem. \\
SciPy capped & Required solve on $\min(C,L)$, filter original $C<L$ & $r\times c$; value is one-sided objective $P$. \\
SciPy dummy columns & Dense or diagonal dummy block costing $L$ & $r\times(c+r)$; value $P$; each dummy has capacity one. \\
SciPy square & Equal endpoint penalties $L/2$ & $(r+c)^2$; value $F=P+(c-r)L/2$. \\
Native rejection & Reusable search terminal of cost $L$ & Original $r\times c$ costs; objective $P$. \\
\midrule
LSAPE O / U & Unchanged LSAPE v4, full buffer & Equal endpoints / one-sided penalty. \\
LSAPE S / T & Specialized code, compact buffer & Ordered rejection scan / guarded cache. \\
LSAPE P & Specialized tracked code, full buffer & U-compatible leading dimension; matched-layout control. \\
\midrule
Campaign boundary & Binary64 inputs, int64 pairs & Public-API checks, conversion, representation, solve and output mapping. \\
Pedestrian boundary & Float32 tracker costs, int32 pairs & Replay also widens captured values to float64 outside timing. \\
LSAPE boundary & Finite, nonnegative binary64 costs; int64 pairs & Validation, conversion, column-major copy, allocation, solve and output. \\
\bottomrule
\end{tabular}

\endgroup
\end{table}

\subsection{Validation, batch controls and memory}

The saved campaign validates feasible outputs and their penalized objectives against the specified reference, checks represented costs and input hashes, and retains rejection cardinalities. Pair-array equality is not required when optimal solutions tie. Diagnostics have their own build and environment identities and are matched to primary input hashes. The saved preflight contains 4,195 passed fixture/method results, 2,864 unsupported combinations and 11 numerical-boundary limitation records per environment, with no blocking failures. The limitations occur in seven padded-layout methods; all 11 records concern a threshold-adjacent fixture or its transpose and have an exact represented-objective gap of $2^{-53}$, although rounded totals agree. Unsupported fixtures and these boundary records are separate from the complete primary timing inventory. Primary input validation uses relative/absolute objective tolerances $10^{-11}$/$10^{-12}$. These checks support the sampled cases, not a proof of all floating-point transitions.

Batch contrasts distinguish Python loops, one-worker batch dispatch, four-worker execution, and controlled workspace lifetimes. Ratios are formed within each workload and environment; absolute batch times are retained separately. Repeated workspaces are tested directly rather than inferred from a single-call comparison. Independent problems must already be available together for inter-problem parallelism to apply.

Every memory condition has five fresh-process trials with resident input before baseline and a 50 ms output hold. Peak capture precedes reference validation. Linux records high-water RSS and Windows peak working set; both retain pre-call headroom below the process's previous peak. Additional peak, final resident growth and exact buffer arithmetic are different quantities. A zero additional high-water increment does not imply zero allocation. Memory orientation inputs are distribution-matched, whereas the timing orientation pairs use exactly transposed values.

\subsection{Application and LSAPE contracts}

\paragraph{Pedestrians.}
The application suite reuses supplied 999-frame excerpts of MOT16-04 and MOT20-03 \cite{milan2016mot16,dendorfer2020mot20}, with supplied ground-truth boxes or cached RF-DETR Nano detections \cite{rfdetr2025software}. The detector cache used RF-DETR 1.10.1 at 768-pixel resolution and confidence 0.2. Centroid and SORT \cite{bewley2016sort} come from pyppbox 4.0.0 \cite{pyppbox2026,siv2022posetreid}, pinned to source commit \texttt{f3f1e8e671d7}. Dataset/cache modifications and third-party notices are recorded in \path{benchmark/application/ATTRIBUTION.md}. Three rounds measure tracker updates and replay of the common captured SciPy matrices. Replay retains the float32 values and an exactly widened float64 view; it measures assignment calls separately from tracker updates. Detector inference, decoding and rendering are outside both boundaries. All eight APIs, including three distinct JVX return interfaces, remain in the supplement. The stock Centroid distance gate and SORT IoU post-filter and unique-feasible bypass are preserved; finite-limit assignment is not substituted into the trackers. The adapter normalizes application output to int32 pairs, unlike the controlled campaign's int64 pairs. Accuracy uses supplied-box motmetrics 1.4.0 \cite{motmetrics2026} evaluation at IoU $\geq0.5$, without the official dataset class/ignore policy. Accuracy and prediction checks are reported separately from objective checks.

\paragraph{Microscopy.}
A separate session on 13 November 2026 uses ByoTrack 2.0.8's Kalman linker~\cite{byotrack2024kalman}, Pylapy 1.2.0 and LAPX 0.13.1 on Windows 11 with Python 3.14.10, SciPy 1.17.1 and CPU-only PyTorch 2.14.1. It runs on the same processor model, logical CPU 0, with one-thread numerical libraries and the Balanced power scheme. Eight excerpts use the final up to 100 imperfect segmentation masks from sequence 02 of the Cell Tracking Challenge~\cite{maska2023ctc}; Hydra uses prepared wavelet detections from zero-based frames 130--229~\cite{hanson2024hydra}. The excerpt selection and preparation recipe follow Reme's earlier MIT-licensed benchmark~\cite{reme2026cellbenchmark}; this session adds the capped control, balanced timing orders and stronger recorded checks, using its own measurements.

All arms use dense \texttt{OPT\_ALT}, shared estimated motion parameters, no optical flow or splitting/merging, and float32 negative log-likelihood costs with finite rejection limits. Stock means the SciPy backend forced through Pylapy's standard $(r+c)$-square expansion, not ByoTrack's default sparse mode or Pylapy's automatic backend selection. Capped SciPy retains the original dimensions and the common adapter; LAPX's \texttt{lapjvr}, selected through Pylapy's \texttt{laprust} backend, handles rejection internally. One warm-up per arm precedes six rounds covering all six arm orders, shuffled with seed 20261113. The whole-linker boundary includes final smoothing and track collection; input preparation and parameter estimation are excluded. Garbage collection is disabled during timing. Reported linker times are medians of six whole-clip means per frame; ratios divide these medians. Separate instrumented runs time association including the adapter, solver and pair conversion, rather than the solver kernel alone. Appendix~\ref{app:microscopy} specifies this boundary and the saved validation.

\paragraph{LSAPE.}
The Windows-only LSAPE study uses the July-2017 LSAPE-v4 Hungarian implementation \cite{lsape2017}, with five arms defined in Table~\ref{tab:methods}. Each of three fresh processes per Windows interpreter runs all 67 cases. The five arms share four inputs per case within each process, giving 12 distinct inputs per case across processes. Each process calibrates to a 30 ms target and shuffles the arm order with a seeded generator in each of nine rounds. This is a frozen protocol rather than the assignment campaign's three-round estimator. Each process estimates a paired-round geometric-mean ratio; the reported central value and range are the median and observed minimum/maximum of the three process estimates. These ranges are descriptive, not confidence intervals.

With $h=\mathrm{float64}(L/2)$ and $p=h+h$, equal endpoint costs $(h,h)$ and one-sided costs $(p,0)$ differ by $h(m-n)$; comparisons restore the same full objective. The prespecified primary comparison ($O/P$) contrasts the original implementation with the specialized padded arm. $O/U$ changes formulation within unchanged upstream code, $U/P$ measures specialization on the same full buffer and stride, $S/T$ changes rejection tracking at fixed compact layout, and $P/T$ changes padded versus compact tracked representation. The specialization also includes indexing and parameter-passing changes, so its components are not individually isolated. All arms copy the real costs; compact storage removes only $n+m+1$ doubles. Here, $\mathrm{DBL\_MAX}$ is the largest finite binary64 value. The frozen bridge accepts finite nonnegative binary64 costs and limits bounded by $\mathrm{DBL\_MAX}/[16(n+m+1)]$, with $(n+1)(m+1)\leq\mathrm{INT\_MAX}$, round-to-nearest and gradual underflow. Signed costs, forbidden edges and required assignment are outside this experiment's contract. Guarded rejection caching falls back to the ordered scan outside its stated numerical range. This experiment introduces no JV row reductions and supplies no JV-versus-LSAPE ranking or measured memory benefit.

\FloatBarrier
\section{Results}\label{sec:results}

\FloatBarrier
\subsection{Coverage and native comparisons}

\begin{figure}[!htbp]
\centering
\includegraphics[width=\linewidth,height=.65\textheight,keepaspectratio]{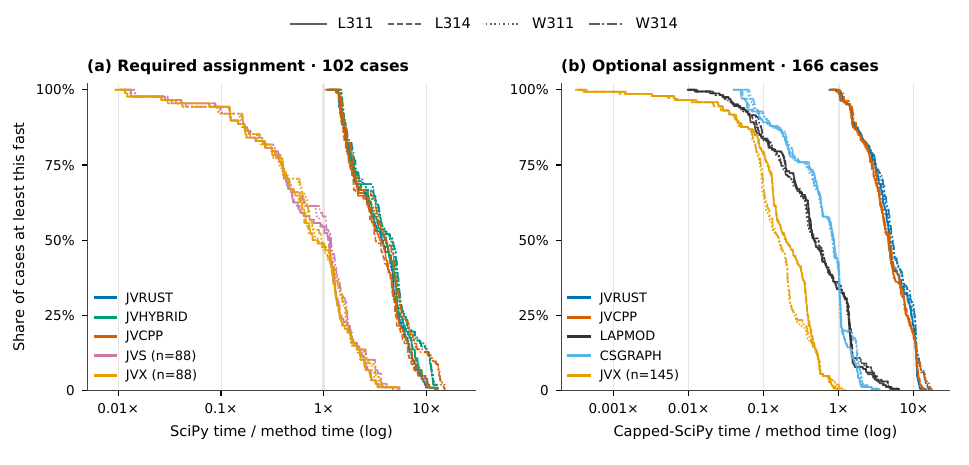}
\caption{Complementary empirical distributions of matching-SciPy/method complete-call ratios. One curve represents one method and environment over that method's planned cases; reduced denominators are stated in the panel legends. Required and optional objectives remain separate. Color identifies method; solid, dashed, dotted and dash-dot lines identify L311, L314, W311 and W314, respectively. The vertical rule is parity and the gray band is $[0.97,1.03]$. Environments are not averaged.}
\label{fig:native}
\end{figure}

Figure~\ref{fig:native} shows the full distribution of matching-SciPy/method ratios for native, legacy dense and sparse solvers on their planned single-matrix cases. Each curve is one solver and environment; its height is the fraction of that solver's admitted cases with a ratio at least as large as the horizontal coordinate. This retains losses as well as the extent of an advantage. All three native methods win all 102 required comparisons under the stated band. Across the four separately summarized environments, JVRUST's median ratio ranges from $3.67$ to $4.66$ for required assignment and $4.72$ to $5.23$ for optional assignment. These are medians over the selected cases within each environment, not pooled estimates. Tables~\ref{tab:nativefamilies} and~\ref{tab:nativeexceptions} give the native family counts and exact non-win workloads.

For optional assignment, JVRUST's wins/ties/losses are $165/1/0$, $165/1/0$, $163/2/1$ and $164/1/1$ in L311/L314/W311/W314. JVCPP's corresponding counts are $163/2/1$, $163/2/1$, $165/1/0$ and $166/0/0$. The denominator is 166 capped-SciPy comparisons in every environment. All optional native ties and losses occur on congested-block inputs in three declared families: formulations, scaling and congestion. The only losses are on the hardest congestion-sweep case, case 041: JVRUST's ratios are approximately $0.984$, $0.985$, $0.916$ and $0.926$; JVCPP's are $0.956$, $0.943$, $1.023$ and $1.036$. The W311 Rust loss corresponds to about 9.2\% extra latency relative to capped SciPy. The other exceptions are practical ties on the optional $256\times320$ formulation and $128\times160$ scaling inputs. Native rejection does not imply universal dominance over an original-dimension formulation.

\FloatBarrier
\subsection{Geometry scaling}

\begin{figure}[!htbp]
\centering
\includegraphics[width=\linewidth,height=.65\textheight,keepaspectratio]{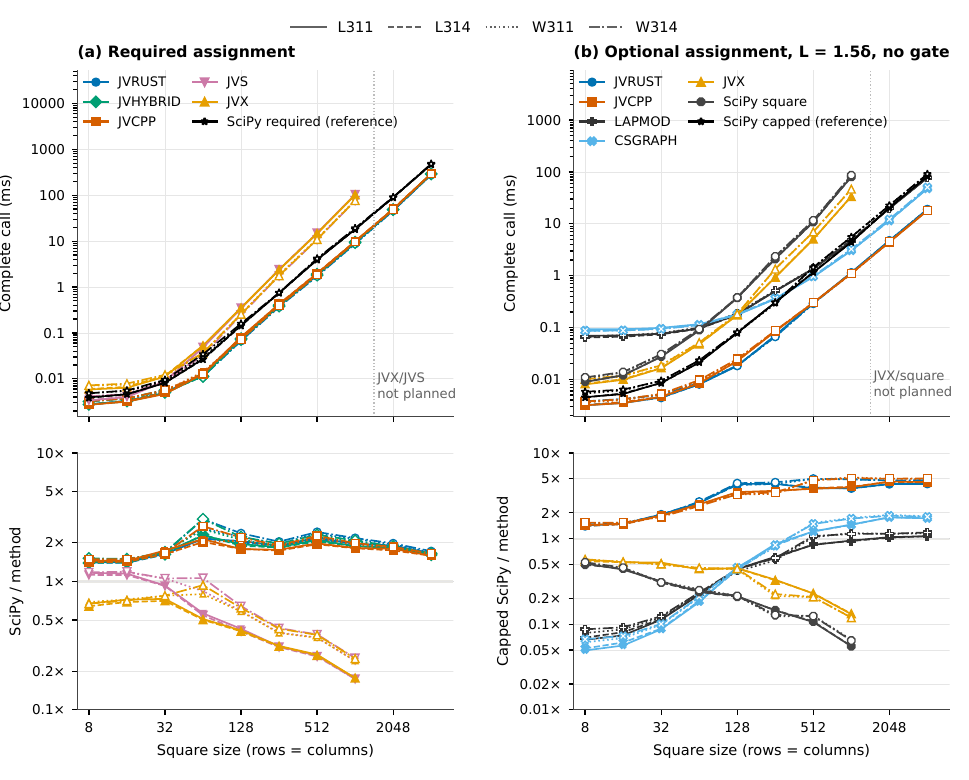}
\caption{Square geometry with $q=0.9$ and motion standard deviation $0.25\delta$: complete-call ms above, matching-SciPy/method ratios below. Required assignment and optional assignment with $L=1.5\delta$ and no gate are separate columns. Method-specific colors and symbols combine with solid/dashed/dotted/dash-dot lines for L311/L314/W311/W314. JVX/JVS are not planned above 1024 rows in the required panel, including unpadded squares; optional JVX and square-expanded SciPy obey the same cutoff. No scaling exponent is fitted.}
\label{fig:geometry}
\end{figure}

Figure~\ref{fig:geometry} links absolute runtime and relative latency over the square geometry size sweep for $q=0.9$ and motion standard deviation $0.25\delta$. Each line retains one method and environment. In this setting, CSGRAPH first exceeds a 3\% advantage over capped SciPy at 512 rows in every environment; every sampled smaller size is a practical loss. The same crossover holds in the $q=0.6$, motion-$\delta$ companion. Across both square settings and all environments, CSGRAPH takes approximately $15$--$20\times$ the capped-SciPy latency at eight rows, including sparse construction. The supplement retains the second setting, $5:4$ geometry and non-geometric scaling with the same line conventions. Appendix~\ref{app:scaling} gives the crossed geometry-sensitivity factors. No full-session observations or extrapolations fill planned exclusions.

\FloatBarrier
\subsection{Representation and padded layout}\label{sec:representations}

\begin{figure}[!htbp]
\centering
\includegraphics[width=\linewidth,height=.65\textheight,keepaspectratio]{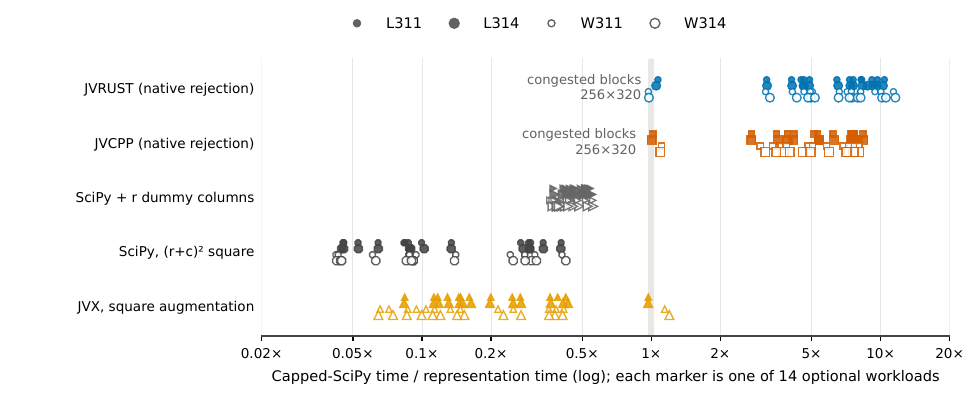}
\caption{Distribution over the 14 optional formulation workloads for each displayed implementation, relative to capped SciPy. One marker is one workload/environment ratio; filled/open markers denote Linux/Windows, and smaller/larger markers denote Python 3.11/3.14. Deterministic offsets separate coincident observations without changing their horizontal values. The labelled congested-block workload retains the native practical ties. The diagonal-dummy and controlled-layout comparisons remain in the supplement.}
\label{fig:representations}
\end{figure}

The equivalent objectives in Section~\ref{sec:formulation} do not require identical storage or work. Figure~\ref{fig:representations} shows five complete-call implementations on all 14 optional formulation workloads; the congested-block input is labelled because it contains the native near-parity exceptions. The supplement retains the diagonal-dummy variant and all 624 controlled layout conditions per environment. An expanded-versus-native gap includes construction and a different solver trajectory; it cannot be attributed entirely to padding allocation. Capped SciPy controls dimensions while retaining different initialization, search and interface work. Constructor-only measurements and buffer dimensions remain secondary evidence.

\FloatBarrier
\subsection{Initialization policies and measured work}\label{sec:policy}

\begin{figure}[!htbp]
\centering
\includegraphics[width=\linewidth,height=.65\textheight,keepaspectratio]{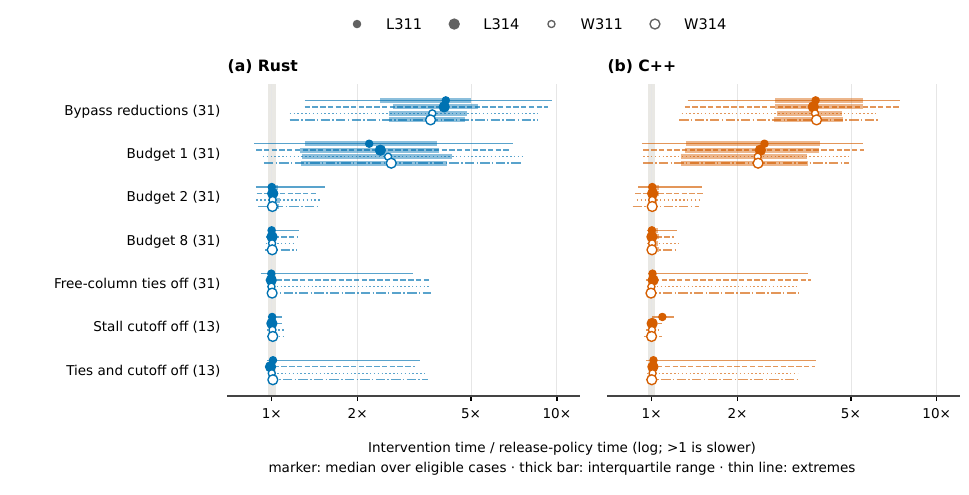}
\caption{Policy distributions within each language and environment. For each intervention, a marker gives the median intervention/budget-4-reference latency ratio over eligible cases; the thick bar is the interquartile range and the thin line the observed minimum--maximum. These are descriptive spreads over different workloads, not confidence intervals. Filled/open markers denote Linux/Windows, and smaller/larger markers denote Python 3.11/3.14. Values above one mean a slower intervention. Table~\ref{tab:counters} uses separately instrumented calls.}
\label{fig:policy}
\end{figure}

Figure~\ref{fig:policy} summarizes the distribution of case-normalized intervention/reference latency ratios for each controlled policy and environment. Median, interquartile range and extremes describe eligible workloads, not sampling uncertainty. Bypassing reductions is approximately $3.6$--$4.1\times$ slower: median intervention/reference latency is $4.077$, $4.024$, $3.655$ and $3.605$ for Rust and $3.760$, $3.689$, $3.742$ and $3.787$ for C++, respectively. Other interventions retain case-level gains and regressions; a favorable median cannot select a universally optimal policy. The supplement retains all eligible per-case responses.

\begin{samepage}
The counters in Table~\ref{tab:counters} show how initialization substitutes for later search. For the optional $256\times320$ base case, bypassing reductions changes mean initializer/search row scans from $512/0$ to $0/700.2$, while augmentations change from 0 to 256. For the tied required case, disabling free-column tie preference increases both initializer and search work. Conversely, disabling the required cutoff in the displayed geometry case increases initializer scans while reducing later search. These are controlled event changes; their relative costs still depend on access patterns and compiled code. Identical event vectors across languages and environments are printed once after verification, rather than averaged.
\par
\end{samepage}

\begin{table}[!htbp]
\centering
\caption{Diagnostic mechanism anchors: mean events per matrix over five frozen inputs, reference $\rightarrow$ intervention. The verified event vectors agree across all four environments and Rust/C++; identical values are shown once. Zeros are literal counts. All six event fields and capacity records remain in the supplement.}
\label{tab:counters}
\begingroup\small
\setlength{\tabcolsep}{3pt}
\renewcommand{\arraystretch}{1.12}
\begin{tabular}{lrrrr}
\toprule
Case; intervention & Init. scans & Search scans & Augmentations & Rejections \\
\midrule
Optional uniform; bypass & 512 $\to$ 0 & 0 $\to$ 700.2 & 0 $\to$ 256 & 0 $\to$ 0 \\
Near square; budget 1 & 482.4 $\to$ 333.2 & 0 $\to$ 399.6 & 0 $\to$ 77.2 & 0 $\to$ 0 \\
Tied base; free ties off & 128 $\to$ 219.8 & 0 $\to$ 91.2 & 0 $\to$ 91 & 0 $\to$ 0 \\
Geometry required; cutoff off & 524.8 $\to$ 597.2 & 5.6 $\to$ 2.6 & 0.4 $\to$ 0.2 & 0 $\to$ 0 \\
Optional blocks; budget 1 & 1285.8 $\to$ 384 & 872.4 $\to$ 1001.6 & 112 $\to$ 128 & 112 $\to$ 112 \\
\bottomrule
\end{tabular}

\endgroup
\end{table}

\FloatBarrier
\subsection{Sparse methods against two references}\label{sec:sparse}

\begin{figure}[!htbp]
\centering
\includegraphics[width=\linewidth,height=.65\textheight,keepaspectratio]{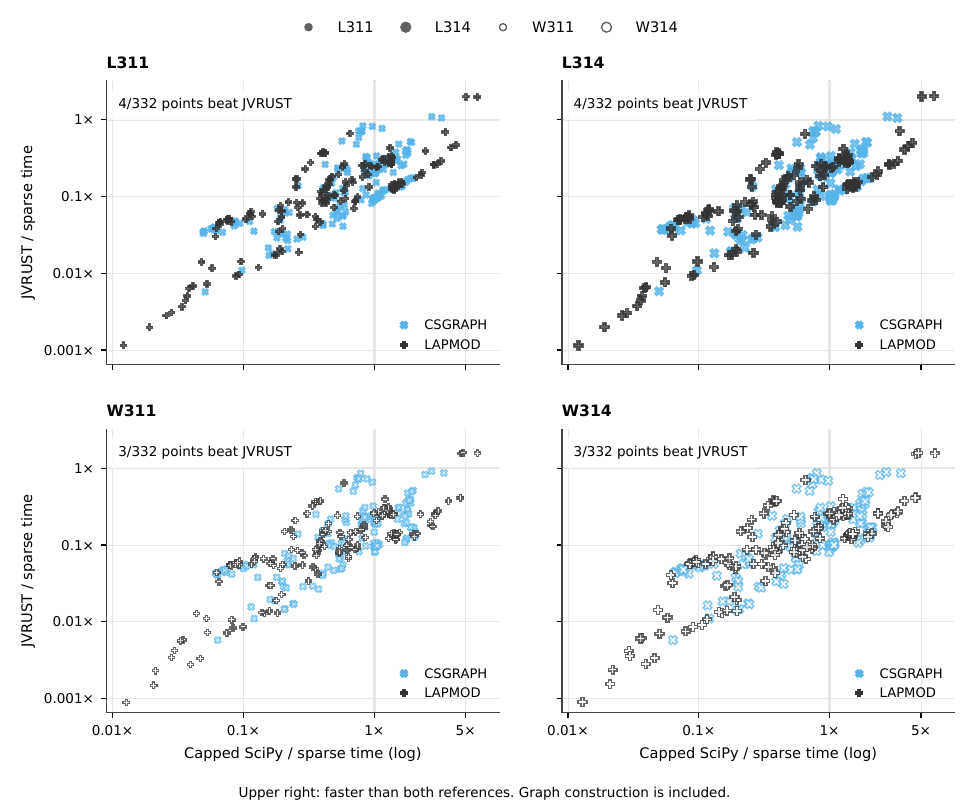}
\caption{All 166 optional cases for each sparse method, with one panel per environment. One point is one case/method: its horizontal coordinate is capped-SciPy/sparse and its vertical coordinate JVRUST/sparse complete-call latency. Method colors distinguish CSGRAPH and LAPMOD. Parity rules and practical-tie bands apply on both log axes. Beating the dense reference does not imply beating JVRUST.}
\label{fig:sparse}
\end{figure}

Beating dense capped SciPy does not imply beating the native implementation. Figure~\ref{fig:sparse} therefore compares each sparse method with both capped SciPy and JVRUST on every optional case. Sparse construction is included, so a solver can benefit from admissible sparsity yet lose once the complete interface is measured. CSGRAPH wins against capped SciPy on 53, 56, 47 and 55 of 166 cases, but against JVRUST on only 2, 2, 0 and 0. LAPMOD's corresponding win counts are 56, 55, 58 and 58 against capped SciPy and 2, 2, 3 and 3 against JVRUST. Retaining both axes exposes cases that a single dense-reference speedup would describe too favorably. Finite-after-gate diagnostics are not substituted for the actual below-limit admissible-edge density.

A post-hoc same-OS comparison identifies seven case/environment events where the primary SciPy reference latency is at least 30\% larger in one Python environment while JVRUST differs by at most 10\%; input hashes agree. Observed reference ratios span 1.404--1.690. Within these events, three CSGRAPH classifications differ between sibling environments. Holding the target CSGRAPH latency fixed and substituting only its sibling reference changes two classifications, not three. Native classifications do not change under either comparison. These are sensitivity checks, not corrected measurements or evidence that reference timing caused all observed differences.

\FloatBarrier
\subsection{Ordered cost structure and rejection}

\begin{figure}[!htbp]
\centering
\includegraphics[width=\linewidth,height=.65\textheight,keepaspectratio]{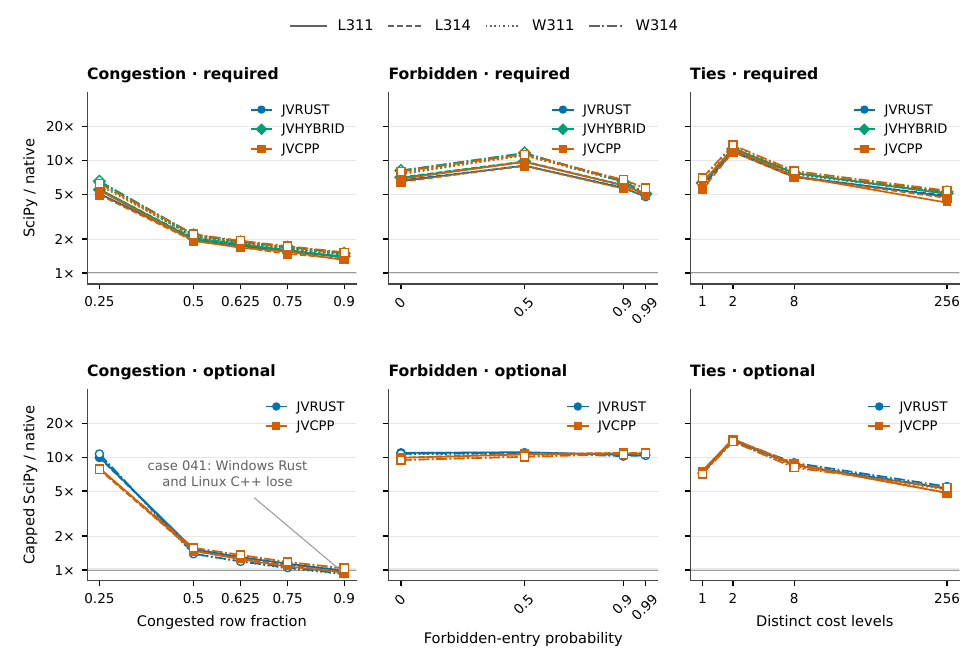}
\caption{Ordered native-solver sweeps at $256\times320$. Columns vary congested row fraction, forbidden-entry probability and quantization levels; rows separate required and optional assignment. One curve is one method/environment, with all other case parameters fixed. Color identifies method; solid/dashed/dotted/dash-dot lines identify L311/L314/W311/W314. Ratios above one favor the native method over required or capped SciPy. The case-041 annotation retains both Rust/Windows and C++/Linux losses.}
\label{fig:structure}
\end{figure}

Figure~\ref{fig:structure} shows the ordered congestion, forbidden-entry and quantization sweeps. Required feasibility is planted in the forbidden-entry sweep. Changing the congested row fraction changes measured latency at fixed dimensions; the sweep has no counters separating initialization from completion work. Its 90\% congestion endpoint, case 041, contains the only optional native losses: Windows JVRUST and Linux JVCPP. The other congested-block exceptions occur in the formulation and scaling families (Table~\ref{tab:nativeexceptions}). The quantized-cost sweep measures sensitivity to tie density; the separate policy intervention and counters test free-column preference.

The rejection-limit appendix links complete-call latency to realized unmatched-row fractions read from the saved validation records. A limit changes the objective and search prices, so this is not a causal experiment on rejection fraction alone. In particular, equal realized cardinality need not imply equal search work or latency.

\FloatBarrier
\subsection{Batching and workspace lifetime}

\begin{figure}[!htbp]
\centering
\includegraphics[width=\linewidth,height=.78\textheight,keepaspectratio]{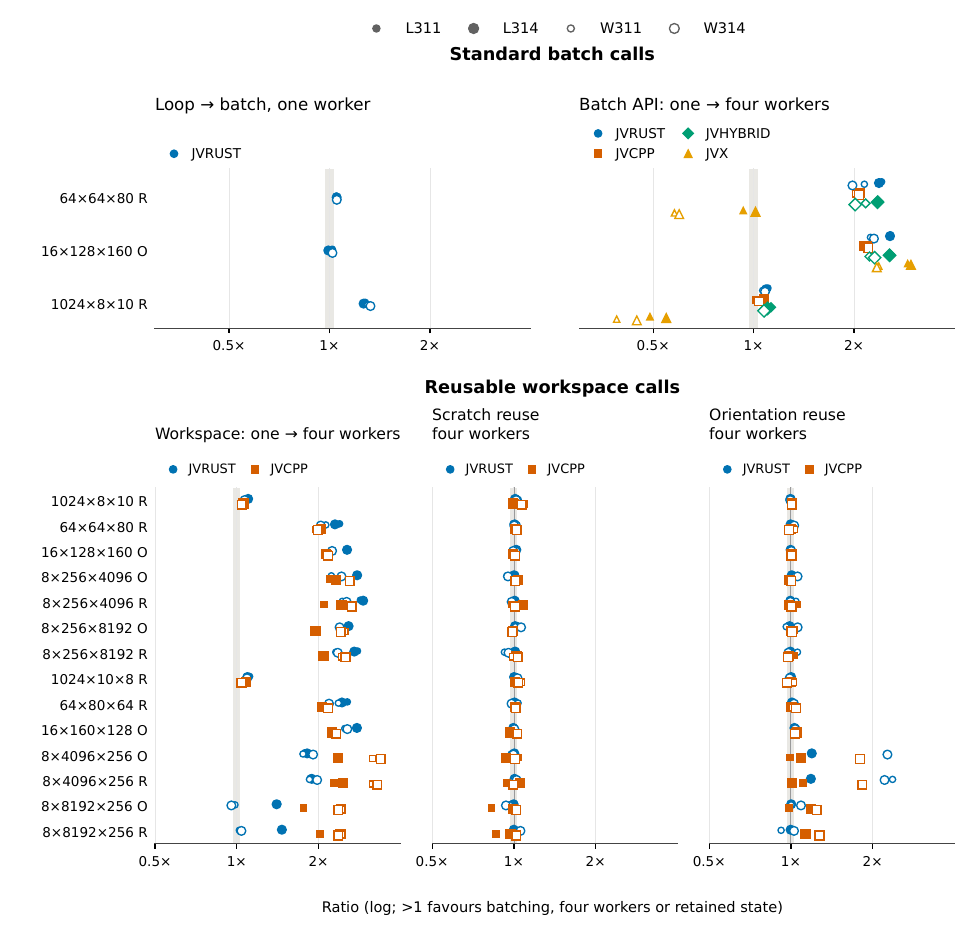}
\caption{Named batch contrasts with workload rows. Each marker is a method/environment ratio: loop/one-worker batch, one/four workers, recreated/retained scratch ($W1/W0$), or recreated/retained orientation storage ($W2/W1$), as labelled. Colors identify methods; filled/open markers denote Linux/Windows, and smaller/larger markers denote Python 3.11/3.14. The workspace panel uses W0 for worker scaling and four workers for reuse. All three standard and 14 workspace workloads remain visible; one-worker reuse and the remaining contrasts are supplementary. Values below one retain regressions.}
\label{fig:batch}
\end{figure}

Figure~\ref{fig:batch} separates loop-to-batch dispatch, worker count and workspace lifetime. A one-worker batch can reduce repeated interface work without introducing parallelism; a four-worker contrast additionally changes scheduling and worker-local workspace use. The workspace controls test retention directly, including effects that can differ for tall inputs. Every declared workload is retained, including near-one results. The supplement also preserves the redundant total-worker-lifetime contrast. Retention is not uniformly beneficial: the L311 C++ four-worker optional $8\times8192\times256$ workload has a recreated/retained solver-state ratio of about 0.828. Worker-count changes can also lose: the tiny JVX batch has one/four-worker ratios of $0.489$, $0.547$, $0.389$ and $0.447$ in L311/L314/W311/W314. The $64\times64\times80$ JVX batch also loses in L311, W311 and W314 (approximately $0.931$, $0.579$ and $0.599$), while L314 is a practical tie. These losses prevent interpreting a batch API or retained workspace as an unconditional speedup. These controlled contrasts support decisions for independent batches but do not promise the same gains for a tracker that presents one matrix at a time.

\FloatBarrier
\subsection{Applications: replay, updates and output differences}\label{sec:applications}

\begin{figure}[!htbp]
\centering
\includegraphics[width=\linewidth,height=.65\textheight,keepaspectratio]{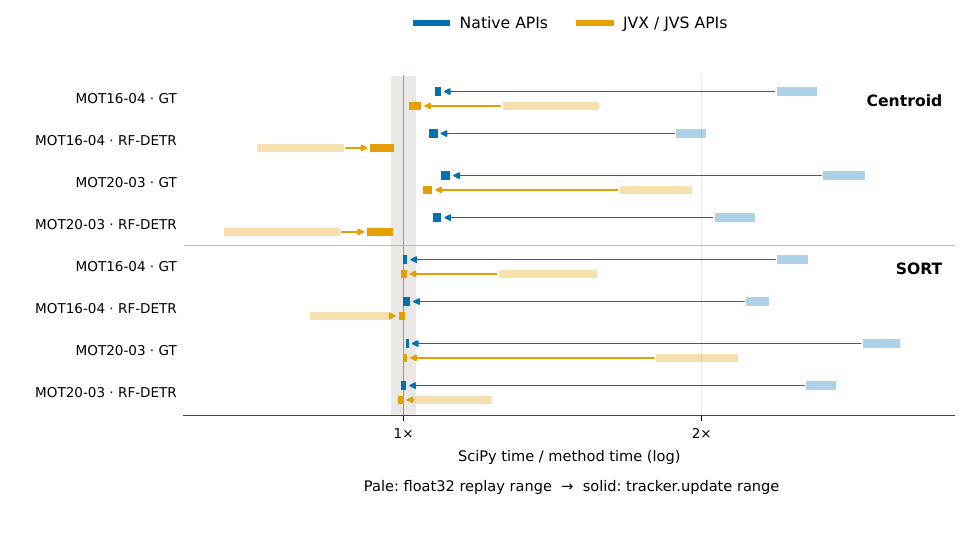}
\caption{Pedestrian tracking with prepared detections on eight pipelines. Each ratio divides SciPy's median time by the corresponding API's median time over three rounds. Values above one favor that API. Pale segments span assignment replay on shared float32 costs and solid segments span tracker updates; arrows connect the nearest range boundaries in that direction. Native ranges contain 12 API/environment observations (three APIs, four environments); legacy ranges contain 16 (three JVX interfaces and JVS, four environments). These envelopes are neither means nor uncertainty intervals. Blue identifies the native group and amber the legacy group. All 224 individual API/pipeline/environment pairs remain in the supplementary table.}
\label{fig:applications}
\end{figure}

\paragraph{Pedestrians.}
Figure~\ref{fig:applications} shows how assignment-replay gains change at the tracker-update boundary within each pipeline. Its native and legacy API groups use observed ranges across their APIs and the four environments, not averaged timings. Each native method has four update wins and four practical ties in each environment: Centroid gains attenuate relative to replay, while SORT updates remain near parity. This measures cached-input tracker updates, not detector-plus-tracker frame rate. Separate SciPy profiles put assignment at 12.5--15.6\% of Centroid update time and 0.9--2.0\% of SORT update time. Replay compares identical captured costs; tracker updates also include motion and bookkeeping. The supplementary per-API table retains every environment-specific replay/update pair, including slower outcomes.

Each environment records 135,702 replay validation checks on 7,539 matrices, covering the eight timed APIs plus the Rust alias and two precision views. All stored nonzero objective-gap counts and mismatch counts are zero. Replay checks and tracking checks answer different questions. Equal objective values on the captured common matrices do not guarantee identical choices under ties, and altered choices can change later tracker inputs. The saved native predictions differ from SciPy on some Centroid pipelines, including both favorable metric changes and additional identity switches. Appendix~\ref{app:applications} retains the workload counts, accuracy deltas and prediction identities; no general accuracy improvement is claimed.

\paragraph{Microscopy.}
Table~\ref{tab:microscopy} retains all nine excerpts and the separate association aggregate. Ratios of six-round median whole-linker times range from 1.03--1.37 for stock SciPy/Rust and 1.02--1.05 for capped SciPy/Rust. Both peak on Hydra at about 725 detections/frame; six of nine capped comparisons fall inside the 3\% band. Across the 54 excerpt-round pairs, Capped/Rust ratios range from 0.935 to 1.165, with a round favoring capped SciPy on three excerpts. Capping therefore narrows the measured whole-linker gap within the same tracker.

The separate association profiles total 1.874, 0.639 and 0.316 seconds for stock SciPy, capped SciPy and Rust over 816 frames; unrounded totals give Stock/Rust and Capped/Rust ratios of 5.929 and 2.023. These include adapter work and pair conversion. On Hydra, capped SciPy and Rust profile means are 4.884 and 2.473\,ms/frame, versus whole-linker medians of 32.998 and 31.571\,ms/frame. The modest overall gain is consistent with most time being spent outside association; the separate runs do not provide an exact time decomposition. All arms have matching saved track digests, and instrumented cost streams and recorded objectives agree. This is saved-record agreement, not an independent revalidation of unavailable matrices or a tracking-accuracy comparison.

\begin{table}[!htbp]
\centering
\caption{Microscopy. Counts give the mean number of detections per frame. Linker times are medians of six whole-clip means with prepared detections. Association totals come from one separate instrumented run per method and excerpt, including the adapter, solver and pair conversion. Ratios use unrounded totals. Stock and Capped denote the two SciPy configurations.}
\label{tab:microscopy}
\begingroup
\small
\setlength{\tabcolsep}{2.5pt}
\begin{tabular}{@{}lrrrrr@{}}
\toprule
\multicolumn{6}{@{}r@{}}{Linker time (ms/frame)} \\
\cmidrule(lr){4-6}
Sequence & Frames & Mean det./frame & Stock & Capped & Rust \\
\midrule
Fluo-C2DL-MSC & 48 & 3.9 & 0.611 & 0.600 & 0.591 \\
PhC-C2DH-U373 & 100 & 6.1 & 0.630 & 0.622 & 0.607 \\
DIC-C2DH-HeLa & 84 & 11.8 & 0.727 & 0.710 & 0.698 \\
BF-C2DL-MuSC & 100 & 18.0 & 0.811 & 0.793 & 0.778 \\
Fluo-N2DH-GOWT1 & 92 & 24.9 & 0.928 & 0.910 & 0.891 \\
BF-C2DL-HSC & 100 & 142.5 & 3.384 & 3.174 & 3.096 \\
Fluo-N2DL-HeLa & 92 & 275.8 & 8.592 & 7.284 & 6.993 \\
PhC-C2DL-PSC & 100 & 342.0 & 11.696 & 9.693 & 9.292 \\
Hydra & 100 & 724.9 & 43.371 & 32.998 & 31.571 \\
\midrule
\multicolumn{3}{@{}l}{Association time, all 816 frames (s)} & 1.874 & 0.639 & 0.316 \\
\multicolumn{3}{@{}l}{Relative to Rust ($\times$)} & 5.929 & 2.023 & 1.000 \\
\bottomrule
\end{tabular}
\endgroup

\end{table}

\FloatBarrier
\subsection{Controlled LSAPE results}\label{sec:lsape-results}

\begin{figure}[!htbp]
\centering
\includegraphics[width=\linewidth,height=.65\textheight,keepaspectratio]{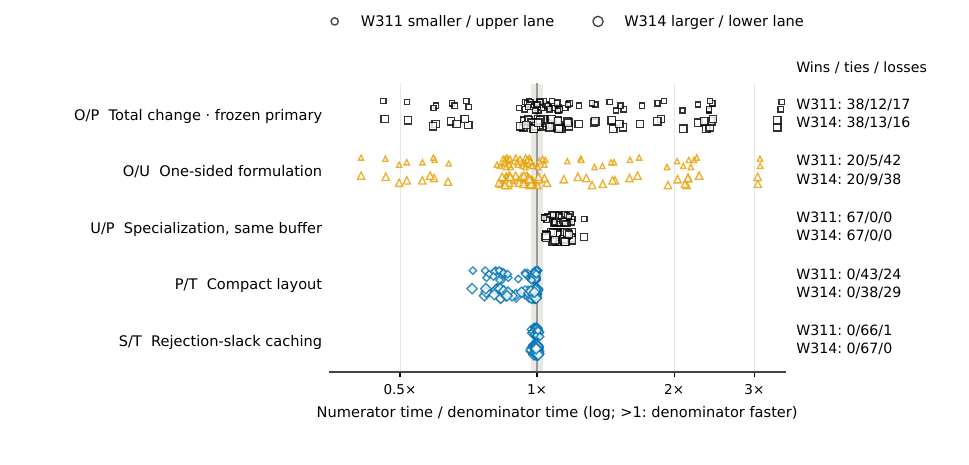}
\caption{All 67 LSAPE cases in each contrast and Windows environment, with the frozen primary $O/P$ first. Each point is the median of three process estimates for one case; smaller/larger open markers denote W311/W314 in separately keyed lanes. Color and shape identify the denominator arm: $P$ black squares, $U$ amber triangles and $T$ blue diamonds. Deterministic vertical spread separates points without changing ratios. Counts beside each row use unrounded values and the practical band. Ratios above one favor the denominator arm. The complete export retains process ranges and all eight saved contrasts.}
\label{fig:lsape}
\end{figure}

Figure~\ref{fig:lsape} shows all 67 case estimates for each of five contrasts, separately for the two Windows interpreters. Table~\ref{tab:lsapefamilies} retains the per-family counts. Formulation ($O/U$), matched-buffer specialization ($U/P$), rejection tracking ($S/T$), and layout ($P/T$) answer distinct questions. Their estimates should not be multiplied from separately summarized medians. Regressions remain as relevant as gains, and a ratio inside the practical band does not establish equivalence. In the primary comparison ($O/P$), the specialized padded arm records 38 wins, 12 ties and 17 losses in W311 and 38 wins, 13 ties and 16 losses in W314. Matched-buffer specialization ($U/P$) wins all 67 cases in each environment. Formulation-only $O/U$ yields wins/ties/losses of $20/5/42$ in W311 and $20/9/38$ in W314. Compact layout $P/T$ yields $0/43/24$ and $0/38/29$; rejection tracking $S/T$ yields $0/66/1$ and $0/67/0$. Compact layout has no improvement exceeding 3\% and is slower beyond the band in 24 and 29 cases, with minimum $P/T$ ratios of approximately $0.723$ and $0.720$. Formulation alone also loses more often than it wins. Rejection tracking is almost entirely within the practical band, with one W311 loss. The specialization advantage therefore does not imply that each transformation helps. The full export also retains absolute arm times and all eight saved comparisons, including total contrasts.

These results characterize the Hungarian adaptation and its measurement boundary. They do not identify the source of the JV/SciPy gap, reproduce the 2016 publication's timings, or establish a general benefit from rejection-slack caching. Process ranges describe the recorded same-host runs rather than independent-day variability.

\FloatBarrier
\subsection{Orientation sensitivity}

\begin{figure}[!htbp]
\centering
\includegraphics[width=\linewidth,height=.65\textheight,keepaspectratio]{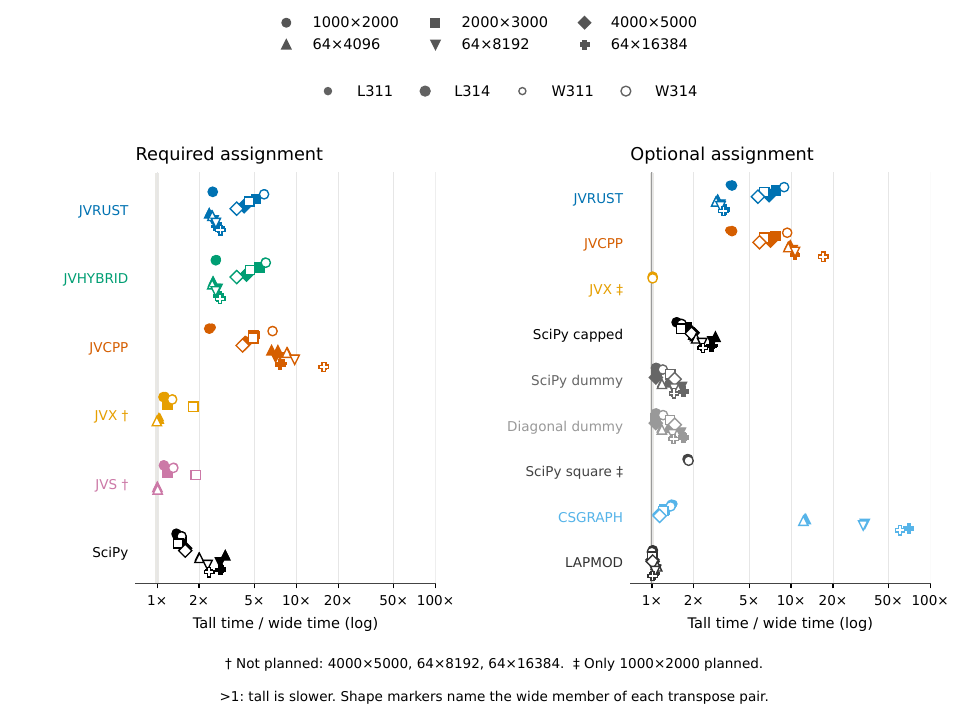}
\caption{Tall/wide complete-call ratios by method, with required and optional objectives separated. One marker is one of six matched shape pairs in one environment; shape symbols distinguish the pairs and colors identify methods. Filled/open markers denote Linux/Windows, and smaller/larger markers denote Python 3.11/3.14. Ratios above one mean a slower tall call. All 74 planned method-pair comparisons per environment are retained, with unavailable shape/method pairs identified as not planned. Wide and tall timings come from distinct attempts with matched underlying values.}
\label{fig:orientation}
\end{figure}

The orientation pairs hold generated values and objective fixed while transposing the original matrix. Figure~\ref{fig:orientation} gives tall/wide complete-call latency, retaining every method planned in both orientations. Ratios are formed from two case attempts within one environment, not from same-attempt paired observations. Input transposition is outside timing, while any internal orientation copy, solve and output handling remain inside it.

For the optional $64\times16384$ pair, JVRUST tall/wide ratios are approximately $3.250$, $3.327$, $3.256$ and $3.266$; JVCPP ratios are $10.681$, $10.647$, $16.827$ and $17.058$ in the standard environment order. CSGRAPH's tall/wide ratios span 60.51--70.54. Its adapter retains the original rows and adds one rejection column per row, so the full matching covers 64 rows of a $64\times16448$ graph for the wide input and 16384 rows of a $16384\times16448$ graph for the tall input. LAPMOD uses a $16448\times16448$ graph in both orientations, and its ratios span 1.011--1.065. These complete-call timings combine representation construction, solver work and output handling.

\FloatBarrier
\subsection{Memory endpoints and counter visibility}

Table~\ref{tab:memory} retains both orientations and every planned method at the extended $256\times4096$ endpoints. Linux has a zero median additional high-water increment in 84 of 91 conditions for each interpreter; Windows has 18 and 17. These zeros cannot rank intrinsic workspace requirements across operating systems. Appendix~\ref{app:memory} compares each raw trial's high-water increment with its net post-call resident growth and retains the baseline headroom that can conceal an increment. Buffer arithmetic remains distinct: optional square expansion needs 144.5 MiB of float64 costs for this 8 MiB original problem, irrespective of the counter's visibility.

\begin{table}[!htbp]
\centering
\caption{Extended memory endpoints in MiB, median [minimum, maximum] over five fresh processes. Linux and Windows counter groups remain separate. A zero high-water increment means no increase above an earlier process peak, not zero allocation. Cost MiB is the arithmetic size of one binary64 representation, not additional allocation; native calls can borrow or internally orient the input. The companion export retains condition-specific headroom. R/O denote required/optional; $0^*$ means no new high-water peak.}
\label{tab:memory}
\begingroup\footnotesize
\setlength{\tabcolsep}{3pt}
\renewcommand{\arraystretch}{1.12}
\begin{tabular}{llrrrrr}
\toprule
 & & & \multicolumn{2}{c}{Linux: \texttt{ru\_maxrss}} & \multicolumn{2}{c}{Windows: peak working set} \\
Shape/objective & Method & Cost MiB & L311 & L314 & W311 & W314 \\
\midrule
256$\times$4096 / R & JVRUST & 8 & \begin{tabular}[t]{@{}r@{}}0*\\{\scriptsize [0,0]}\end{tabular} & \begin{tabular}[t]{@{}r@{}}0*\\{\scriptsize [0,0]}\end{tabular} & \begin{tabular}[t]{@{}r@{}}0*\\{\scriptsize [0,0]}\end{tabular} & \begin{tabular}[t]{@{}r@{}}0*\\{\scriptsize [0,0]}\end{tabular} \\
 & JVCPP & 8 & \begin{tabular}[t]{@{}r@{}}0*\\{\scriptsize [0,0]}\end{tabular} & \begin{tabular}[t]{@{}r@{}}0*\\{\scriptsize [0,0]}\end{tabular} & \begin{tabular}[t]{@{}r@{}}0*\\{\scriptsize [0,0]}\end{tabular} & \begin{tabular}[t]{@{}r@{}}0*\\{\scriptsize [0,0]}\end{tabular} \\
 & SciPy req. & 8 & \begin{tabular}[t]{@{}r@{}}0*\\{\scriptsize [0,0]}\end{tabular} & \begin{tabular}[t]{@{}r@{}}0*\\{\scriptsize [0,0]}\end{tabular} & \begin{tabular}[t]{@{}r@{}}5.17\\{\scriptsize [5.15,5.18]}\end{tabular} & \begin{tabular}[t]{@{}r@{}}5.15\\{\scriptsize [5.14,5.16]}\end{tabular} \\
 & JVX & 128 & \begin{tabular}[t]{@{}r@{}}29.11\\{\scriptsize [29.00,29.54]}\end{tabular} & \begin{tabular}[t]{@{}r@{}}35.98\\{\scriptsize [35.74,36.34]}\end{tabular} & \begin{tabular}[t]{@{}r@{}}125.07\\{\scriptsize [125.03,125.08]}\end{tabular} & \begin{tabular}[t]{@{}r@{}}125.05\\{\scriptsize [125.04,125.05]}\end{tabular} \\
\addlinespace[2pt]
256$\times$4096 / O & JVRUST & 8 & \begin{tabular}[t]{@{}r@{}}0*\\{\scriptsize [0,0]}\end{tabular} & \begin{tabular}[t]{@{}r@{}}0*\\{\scriptsize [0,0]}\end{tabular} & \begin{tabular}[t]{@{}r@{}}0*\\{\scriptsize [0,0]}\end{tabular} & \begin{tabular}[t]{@{}r@{}}0*\\{\scriptsize [0,0]}\end{tabular} \\
 & JVCPP & 8 & \begin{tabular}[t]{@{}r@{}}0*\\{\scriptsize [0,0]}\end{tabular} & \begin{tabular}[t]{@{}r@{}}0*\\{\scriptsize [0,0]}\end{tabular} & \begin{tabular}[t]{@{}r@{}}0*\\{\scriptsize [0,0]}\end{tabular} & \begin{tabular}[t]{@{}r@{}}0*\\{\scriptsize [0,0]}\end{tabular} \\
 & SciPy capped & 8 & \begin{tabular}[t]{@{}r@{}}0*\\{\scriptsize [0,0]}\end{tabular} & \begin{tabular}[t]{@{}r@{}}0*\\{\scriptsize [0,0]}\end{tabular} & \begin{tabular}[t]{@{}r@{}}5.16\\{\scriptsize [5.15,5.18]}\end{tabular} & \begin{tabular}[t]{@{}r@{}}5.16\\{\scriptsize [5.14,5.18]}\end{tabular} \\
 & SciPy dummy & 8.5 & \begin{tabular}[t]{@{}r@{}}0*\\{\scriptsize [0,0]}\end{tabular} & \begin{tabular}[t]{@{}r@{}}0*\\{\scriptsize [0,0]}\end{tabular} & \begin{tabular}[t]{@{}r@{}}5.70\\{\scriptsize [5.68,5.70]}\end{tabular} & \begin{tabular}[t]{@{}r@{}}5.69\\{\scriptsize [5.67,5.73]}\end{tabular} \\
 & SciPy square & 144.5 & \begin{tabular}[t]{@{}r@{}}83.92\\{\scriptsize [83.65,84.09]}\end{tabular} & \begin{tabular}[t]{@{}r@{}}95.77\\{\scriptsize [95.52,95.90]}\end{tabular} & \begin{tabular}[t]{@{}r@{}}141.83\\{\scriptsize [141.81,141.83]}\end{tabular} & \begin{tabular}[t]{@{}r@{}}141.80\\{\scriptsize [141.80,141.82]}\end{tabular} \\
\addlinespace[2pt]
4096$\times$256 / R & JVRUST & 8 & \begin{tabular}[t]{@{}r@{}}0*\\{\scriptsize [0,0]}\end{tabular} & \begin{tabular}[t]{@{}r@{}}0*\\{\scriptsize [0,0]}\end{tabular} & \begin{tabular}[t]{@{}r@{}}5.03\\{\scriptsize [5.03,5.07]}\end{tabular} & \begin{tabular}[t]{@{}r@{}}5.05\\{\scriptsize [5.04,5.09]}\end{tabular} \\
 & JVCPP & 8 & \begin{tabular}[t]{@{}r@{}}0*\\{\scriptsize [0,0]}\end{tabular} & \begin{tabular}[t]{@{}r@{}}0*\\{\scriptsize [0,0]}\end{tabular} & \begin{tabular}[t]{@{}r@{}}5.04\\{\scriptsize [5.04,5.05]}\end{tabular} & \begin{tabular}[t]{@{}r@{}}5.05\\{\scriptsize [5.05,5.18]}\end{tabular} \\
 & SciPy req. & 8 & \begin{tabular}[t]{@{}r@{}}0*\\{\scriptsize [0,0]}\end{tabular} & \begin{tabular}[t]{@{}r@{}}0*\\{\scriptsize [0,0]}\end{tabular} & \begin{tabular}[t]{@{}r@{}}13.18\\{\scriptsize [13.16,13.19]}\end{tabular} & \begin{tabular}[t]{@{}r@{}}13.16\\{\scriptsize [13.14,13.19]}\end{tabular} \\
 & JVX & 128 & \begin{tabular}[t]{@{}r@{}}28.52\\{\scriptsize [28.45,28.83]}\end{tabular} & \begin{tabular}[t]{@{}r@{}}35.86\\{\scriptsize [35.78,36.27]}\end{tabular} & \begin{tabular}[t]{@{}r@{}}125.03\\{\scriptsize [125.03,125.14]}\end{tabular} & \begin{tabular}[t]{@{}r@{}}125.05\\{\scriptsize [125.04,125.13]}\end{tabular} \\
\addlinespace[2pt]
4096$\times$256 / O & JVRUST & 8 & \begin{tabular}[t]{@{}r@{}}0*\\{\scriptsize [0,0]}\end{tabular} & \begin{tabular}[t]{@{}r@{}}0*\\{\scriptsize [0,0]}\end{tabular} & \begin{tabular}[t]{@{}r@{}}5.07\\{\scriptsize [5.07,5.07]}\end{tabular} & \begin{tabular}[t]{@{}r@{}}5.09\\{\scriptsize [5.08,5.18]}\end{tabular} \\
 & JVCPP & 8 & \begin{tabular}[t]{@{}r@{}}0*\\{\scriptsize [0,0]}\end{tabular} & \begin{tabular}[t]{@{}r@{}}0*\\{\scriptsize [0,0]}\end{tabular} & \begin{tabular}[t]{@{}r@{}}5.07\\{\scriptsize [5.07,5.11]}\end{tabular} & \begin{tabular}[t]{@{}r@{}}5.08\\{\scriptsize [5.08,5.21]}\end{tabular} \\
 & SciPy capped & 8 & \begin{tabular}[t]{@{}r@{}}0*\\{\scriptsize [0,0]}\end{tabular} & \begin{tabular}[t]{@{}r@{}}0*\\{\scriptsize [0,0]}\end{tabular} & \begin{tabular}[t]{@{}r@{}}13.18\\{\scriptsize [13.16,13.19]}\end{tabular} & \begin{tabular}[t]{@{}r@{}}13.17\\{\scriptsize [13.15,13.19]}\end{tabular} \\
 & SciPy dummy & 8.5 & \begin{tabular}[t]{@{}r@{}}0*\\{\scriptsize [0,0]}\end{tabular} & \begin{tabular}[t]{@{}r@{}}0*\\{\scriptsize [0,0]}\end{tabular} & \begin{tabular}[t]{@{}r@{}}5.70\\{\scriptsize [5.70,5.70]}\end{tabular} & \begin{tabular}[t]{@{}r@{}}5.69\\{\scriptsize [5.66,5.70]}\end{tabular} \\
 & SciPy square & 144.5 & \begin{tabular}[t]{@{}r@{}}81.94\\{\scriptsize [81.22,82.43]}\end{tabular} & \begin{tabular}[t]{@{}r@{}}94.71\\{\scriptsize [94.31,95.63]}\end{tabular} & \begin{tabular}[t]{@{}r@{}}141.84\\{\scriptsize [141.82,141.84]}\end{tabular} & \begin{tabular}[t]{@{}r@{}}141.79\\{\scriptsize [141.77,141.82]}\end{tabular} \\
\addlinespace[2pt]
\bottomrule
\end{tabular}

\endgroup
\end{table}
\FloatBarrier
\section{Discussion}

The native implementations combine established initialized rectangular JV and uniform-rejection ideas in usable interfaces. Their required and optional results support a practical advantage over the specified compact SciPy references on most measured cases, with congested-block exceptions retained. The capped reference is essential: comparing only with square expansion would conflate avoiding dummy costs with the remaining implementation differences.

The controls make the interpretation more specific. Representation and layout panels separate alternative buffers; policy variants change individual initialization choices within matched builds; counters expose a shift between initializer and completion work; batch controls distinguish dispatch, workers and retention; and orientation pairs reveal different complete-call responses to the same costs. None of these comparisons assigns the entire native/SciPy difference to one component. The sparse comparisons also show why a favorable ratio against one baseline is insufficient to establish a solver ranking.

JVRUST and JVHYBRID share the rectangular Rust core, so differences there cannot establish a different rectangular algorithm. On required squares, the laprust and LAPX timings differ in both completion kernel and package build, so they cannot isolate the cost of completion. The JVCPP comparison also measures a complete implementation. Across environments, matched package versions and verified placement improve environmental control without turning separate operating-system and interpreter sessions into causal experiments.

For applications, the assignment objective, tie convention, represented precision and output interface must match the intended contract. Replay isolates common matrix calls but is not the evolving tracker workload after a different matching decision. Tracker updates and the microscopy whole linker include work beyond assignment, and detector inference adds another boundary. The microscopy association profiles retain a larger capped SciPy/Rust ratio than the whole linker, but they are separate instrumented runs and do not isolate a pure solver-kernel speedup. The observed attenuation and prediction differences justify benchmarking the actual pipeline rather than extrapolating an isolated solver speedup.

\FloatBarrier
\section{Limitations and threats to validity}

There is one primary session per environment, on one recorded processor model. The study does not estimate variation across days, machines or compilers. Frequency, power-management behavior and background activity remain uncontrolled. Three rounds over five frozen inputs give descriptive within-session estimates, not population confidence intervals or significance tests. Within-attempt rounds cannot detect a persistent attempt-level shift. The seven post-hoc reference discrepancies in Section~\ref{sec:sparse} show why reference stability matters, but interpreter, build and collection session change together; the data cannot identify a causal attempt effect. All primary observations and classifications are retained. The practical tie band is a reporting convention, and a tie does not prove numerical or performance equivalence.

The initialization heuristic was selected using related workload families. Controlled ablations characterize the selected panel, but do not establish a held-out optimum or universal policy. The synthetic geometry is tracking-like rather than a sample from the application corpus. The light profile deliberately omits two geometry settings outside the complete panels and stops selected baselines above 1024 rows. No missing observation is extrapolated. The separate full-session sensitivity check supports a bounded reporting choice, not statistical equivalence of protocols.

The proofs hold in exact arithmetic. Float64 working state cannot recover distinctions lost during cost conversion and does not prove arbitrary-range floating-point optimality. As a worked example outside the frozen preflight, take \(L=1\) and the binary64 matrix \(C=\left(\begin{smallmatrix}0&1/2\\1/2-2^{-54}&1\end{smallmatrix}\right)\). JVRUST, JVHYBRID, JVCPP and the capped SciPy adapter return only the zero-cost edge, giving a penalized objective of \(1\). The two off-diagonal edges attain the exact minimum \(1-2^{-54}\), which also rounds to \(1\) in binary64. At exact ties, pair identities and cardinality can also differ. Saved validation supports the actual sampled inputs and represented objectives, not every code transition or error path.

The pedestrian applications cover two cached clips, two detection sources and two trackers. The microscopy study adds nine selected excerpts in one separate Windows session; six timing rounds characterize within-session variation rather than independent replication. Its validation rests on saved track/cost digests and objective gaps, and its package-version record lacks a native-wheel binary hash. Detector preparation is input provenance, not a newly measured stage. Replay gains and tracker-update gains cannot establish a general full-pipeline speedup, and mixed tracking metric changes cannot establish an accuracy benefit. The frozen LSAPE contract is narrower than the JV interface and its process replications remain Windows-only. It is not a direct comparison with JV or every compact error-correcting algorithm.

Memory counters have platform-specific semantics and prior-peak headroom. Neither a zero increment nor final resident growth measures total allocation. The trials do not establish cross-platform memory efficiency or long-running pool retention. Constructor timings, instrumented events and primary complete-call timings have different boundaries and must not be interchanged.

\FloatBarrier
\section{Conclusion}

Native rectangular rejection integrates established assignment principles without requiring square expansion. In each of four environments, the measured Rust, hybrid and C++ implementations win all 102 required comparisons against rectangular SciPy under a 3\% practical band. JVRUST's median reference/method ratio ranges from $3.67$ to $4.66$ for required assignment and $4.72$ to $5.23$ for optional assignment, with each environment summarized separately. Optional gains remain workload-dependent: congested-block inputs produce ties and losses, including 9.2\% extra JVRUST latency in the worst measured Windows comparison with capped SciPy. The controls show why representation, initialization, input structure, orientation and interface work all matter when selecting an implementation. Pedestrian replay gains attenuate in tracker updates. In microscopy, capped SciPy approaches Rust in whole-linker time while separate association profiles retain a larger gap. The LSAPE study remains a separate within-Hungarian experiment. The saved-data figures and indexed evidence support these descriptive results without implying algorithmic priority, universal superiority or unmeasured application benefits.

\FloatBarrier
\section*{Data and software availability}

The project repository is \href{https://github.com/rathaROG/lap-paper}{github.com/rathaROG/lap-paper}. The Cell Tracking Challenge organizers granted permission to publish the microscopy benchmark results reported here.

The four primary \texttt{paper-run-light} sessions under \path{benchmark/runs/} are \path{linux-py3.11.18-20261031-0455}, \path{linux-py3.14.10-20261031-0630}, \path{windows-py3.11.18-20261030-2322} and \path{windows-py3.14.10-20261031-0322}. The manifest in \path{benchmark/evidence/} identifies each archived file, checksum and source root. Four archives contain the primary observations and raw timing and validation logs. The fifth contains the full session \path{linux-py3.11.18-20261030-1434/paper-run} used only for protocol sensitivity. Native module hashes and build identifiers distinguish release, policy and diagnostic variants.

Install \path{benchmark/requirements-artifacts.txt}. From the repository root, run \texttt{python} \path{benchmark/paper_evidence.py} \texttt{extract} to restore and verify the saved files, then \texttt{python} \path{benchmark/generate_paper_artifacts.py} to regenerate the figures, tables and indexed supplement in \path{latex/images/current/}, \path{latex/generated/} and \path{benchmark/paper-artifacts/}. This uses saved observations without executing a solver. Compiling \path{latex/supplementary_figures.tex} produces \path{output/pdf/supplementary_figures.pdf}. The November driver, observations, checksums, attribution and MIT notice are in \path{benchmark/cell_tracking/}. The same generator produces its microscopy tables and S7 exports. These exports omit the source images, masks, detections, tracks and cost matrices.

\FloatBarrier
\section*{Acknowledgements}

The implementations retain attribution to Tomas Kazmar's lap, Vadim Markovtsev's src-d/lapjv, and the contributors acknowledged in the source distributions. The assignment and error-correction literature provides the theoretical background and comparison points for this implementation study.

\FloatBarrier
\section*{AI assistance disclosure}

The first author led the research and proposed the approach studied in this paper. AI coding assistants (OpenAI Codex and Anthropic Claude Code) assisted under the authors' directions with source inspection, experiment tooling, evidence checks, figure generation and manuscript drafting. Saved artifacts identify the computations supporting the results. Responsibility for reviewing the scientific claims, attribution and final text remains with the authors.

\FloatBarrier
\renewcommand{\bibfont}{\small}
\setlength{\bibsep}{0pt plus 1pt}
\bibliographystyle{unsrtnat}
\bibliography{references}

\FloatBarrier
\clearpage
\appendix
\section{Scaling and geometric sensitivity}\label{app:scaling}

The supplementary figures use family-wise line panels for the 79 non-geometric scaling, large-square, large-rectangle and extreme-rectangle cases, alongside the $5:4$ and second-setting geometry companions. Each curve holds objective, generator, orientation, limit policy, method and environment fixed apart from its declared sweep variable. Per-case policy responses and controlled layouts remain separate panels. Exact values for all 268 single-matrix cases remain in the indexed exports.

\begin{figure}[!htbp]
\centering
\includegraphics[width=\linewidth,height=.68\textheight,keepaspectratio]{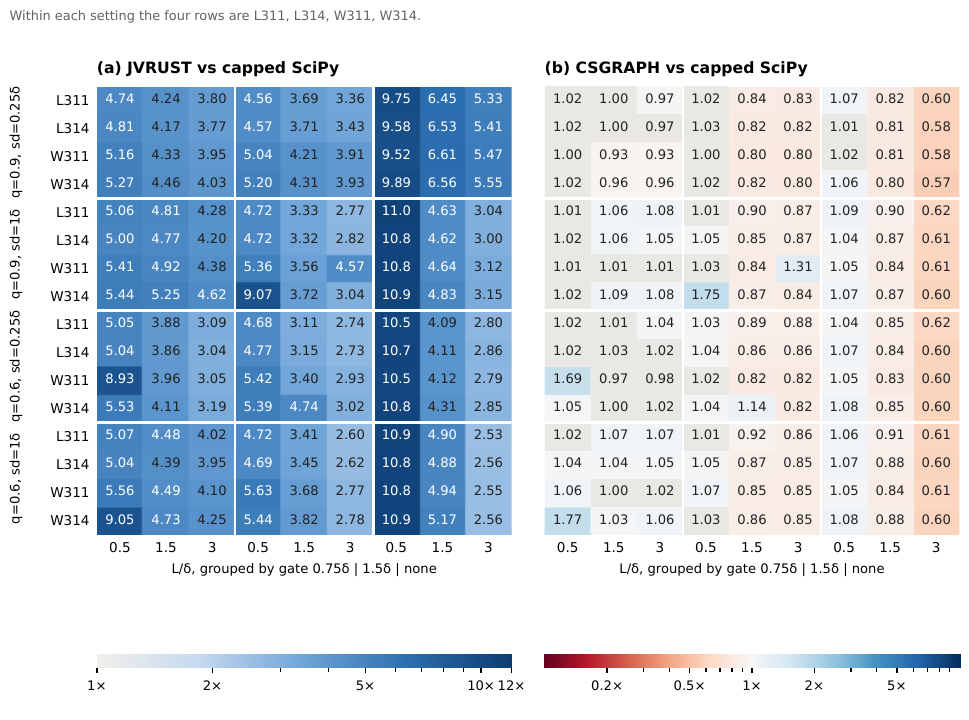}
\caption{Geometry sensitivity at $256\times320$: JVRUST and CSGRAPH relative to capped SciPy. Columns cross gate and $L/\delta$; row groups cross survival and motion, retaining L311/L314/W311/W314 separately within each group. Each printed cell is one observed complete-call ratio. The color key is logarithmic, and classification uses unrounded values. Repeated or reference-sensitive cells remain recorded observations, not averaged or corrected measurements.}
\label{fig:sensitivity}
\end{figure}

\FloatBarrier
Figure~\ref{fig:sensitivity} uses a heatmap for the crossed $256\times320$ geometry factors: four survival/motion settings, three gates and three limits, with environments as separate rows. Cells use capped-SciPy/method ratios. This is the only heatmap; it shows the crossed-factor comparisons directly rather than joining different settings as a size trend.

\FloatBarrier
\section{Rejection-limit sweep}

\begin{figure}[!htbp]
\centering
\includegraphics[width=\linewidth,height=.68\textheight,keepaspectratio]{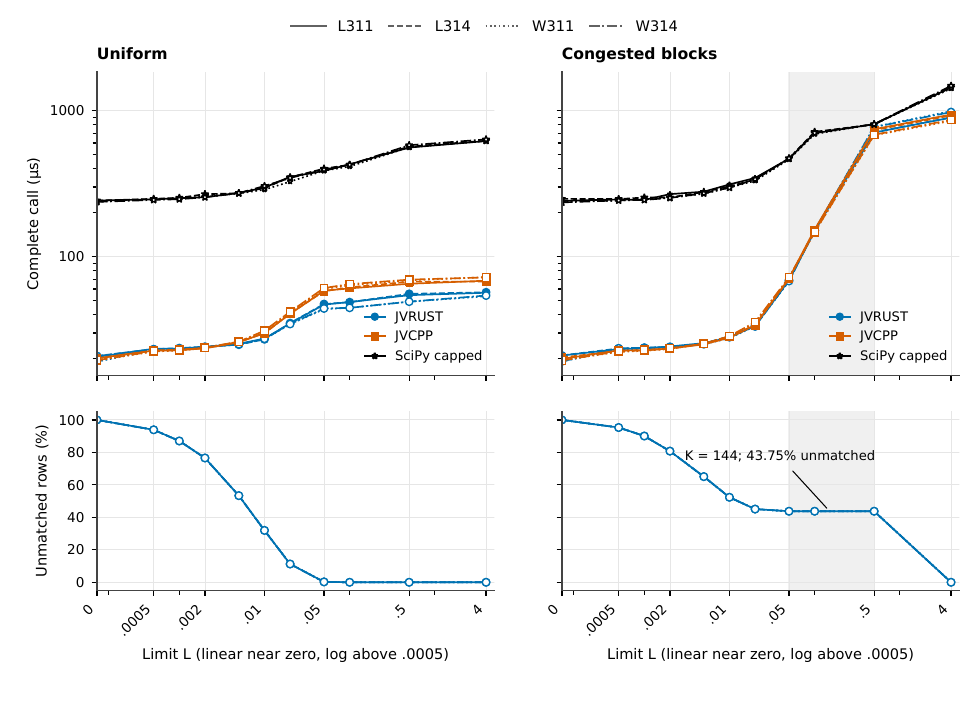}
\caption{Linked rejection-limit sweeps for uniform and congested-block inputs. Runtime panels retain complete-call latency separately by method and environment; companion panels show the smaller-side unmatched fraction derived from the five saved input cardinalities. Solid/dashed/dotted/dash-dot lines identify L311/L314/W311/W314. The common limit axis is linear from zero through 0.0005 and logarithmic above it. Constant cardinality does not imply constant latency; no solver or input generator was rerun to obtain these values.}
\label{fig:rejection}
\end{figure}

The measured limit sweep in Figure~\ref{fig:rejection} reuses saved costs within each family and reads realized cardinality from validation records. No matrix is regenerated and no solver is rerun to obtain the plotted unmatched fractions. In the congested-block $256\times320$ case, every saved input has 144 matches at both $L=0.05$ and $L=0.5$, leaving 43.75\% of smaller-side rows unmatched. JVRUST latency nevertheless changes from 68.94 to 706.63 $\mu$s in L311, 68.91 to 710.73 $\mu$s in L314, 67.93 to 773.42 $\mu$s in W311, and 68.08 to 771.40 $\mu$s in W314. These within-environment changes exceed tenfold at identical cardinality; they do not isolate the rejected fraction as the cause.

\FloatBarrier

\FloatBarrier
\section{Memory-counter visibility}\label{app:memory}

Each panel retains all 455 fresh-process trials from its environment. Zero high-water increments occur in 420, 420, 88 and 82 trials in L311/L314/W311/W314. In each Linux environment, 80 trials have zero high-water increment despite more than 1 MiB of post-call resident growth; no Windows trial has that combination. This describes visibility under the recorded headroom, not intrinsic solver memory efficiency. The complete memory export retains trial-level counter names, pre/post RSS, peaks, headroom and condition identities.

\begin{figure}[!htbp]
\centering
\includegraphics[width=\linewidth,height=.73\textheight,keepaspectratio]{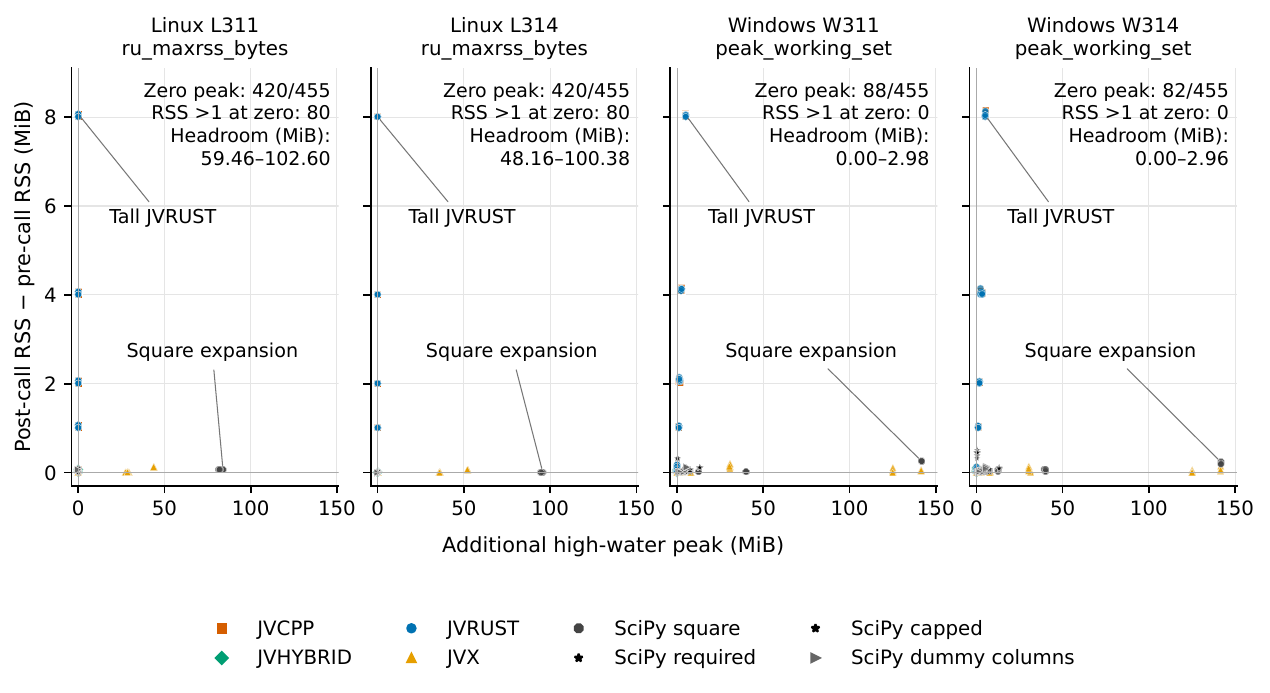}
\caption{Additional high-water increment versus net post-call resident growth, in MiB, with true zeros retained on linear axes. Each panel identifies its operating-system counter and baseline headroom. Net resident growth is a snapshot difference, not a peak or allocation count.}
\label{fig:memoryvisibility}
\end{figure}

\FloatBarrier
\section{Application workload and accuracy}\label{app:applications}

The workload table counts each frozen pedestrian pipeline once, rather than once per API or environment. Replay uses the common matrices captured through SciPy. Tracker updates can evolve different assignments and predictions; the accuracy table therefore keeps solver identities and differences explicit. All precision views, absolute times, stored speedups, round values, validation details and secondary instrumented solver fractions remain in the application supplement. Native Centroid IDF1 changes by $+0.524$ percentage points on MOT16 ground-truth boxes and $-0.008$ on MOT20 RF detections, with no change on the other two pipelines. The JVX interfaces and JVS change Centroid IDF1 by $+0.333$ on MOT16 RF and $-0.397$ on MOT20 RF. SORT metric values agree; prediction identities and identity-switch changes remain separately reported in Table~\ref{tab:accuracy}.

\begin{table}[!htbp]
\centering
\caption{Frozen application workload inventory. Frames, captured matrices and source identities describe the supplied corpus, not a wider tracking population.}
\label{tab:appworkload}
\begingroup\small
\setlength{\tabcolsep}{3pt}
\renewcommand{\arraystretch}{1.12}
\begin{tabular}{llrrrr}
\toprule
Pipeline & Tracker & Frames & Matrices & Rows & Columns \\
\midrule
MOT16-04 / GT & Centroid & 999 & 998 & 47--56 & 47--56 \\
MOT16-04 / GT & Sort & 999 & 998 & 47--56 & 47--56 \\
MOT16-04 / RF-DETR & Centroid & 999 & 998 & 22--39 & 23--39 \\
MOT16-04 / RF-DETR & Sort & 999 & 976 & 23--39 & 24--42 \\
MOT20-03 / GT & Centroid & 999 & 998 & 84--134 & 84--134 \\
MOT20-03 / GT & Sort & 999 & 998 & 84--134 & 84--134 \\
MOT20-03 / RF-DETR & Centroid & 999 & 998 & 34--62 & 34--62 \\
MOT20-03 / RF-DETR & Sort & 999 & 575 & 36--62 & 39--66 \\
\bottomrule
\end{tabular}

\endgroup
\end{table}

\begin{table}[!htbp]
\centering
\caption{Supplied-box tracking validation and metric differences relative to SciPy. Percentage-point changes, switch counts and prediction differences are distinct from replay objective agreement. All retained differences are descriptive and do not establish a general accuracy benefit. Native denotes JVRUST/JVHYBRID/JVCPP; C/S denote Centroid/SORT, M16/M20 the two clips, GT/RF the detection source. The metric vectors and prediction identities agree across environments, so these identical results are printed once.}
\label{tab:accuracy}
\begingroup\footnotesize
\setlength{\tabcolsep}{3pt}
\renewcommand{\arraystretch}{1.12}
\begin{tabular}{llrrrrrl}
\toprule
Pipeline & Solver group & $\Delta$IDF1 & $\Delta$MOTA & $\Delta$IDs & $\Delta$FP & $\Delta$FN & Predictions \\
\midrule
M16 / gt / C & Native & +0.524 & +0.002 & -1 & +0 & +0 & Different \\
M16 / gt / C & JVX APIs/JVS & +0.000 & +0.000 & +0 & +0 & +0 & Different \\
M16 / gt / S & Native & +0.000 & +0.000 & +0 & +0 & +0 & Same \\
M16 / gt / S & JVX APIs/JVS & +0.000 & +0.000 & +0 & +0 & +0 & Same \\
M16 / RF / C & Native & +0.000 & -0.006 & +3 & +0 & +0 & Different \\
M16 / RF / C & JVX APIs/JVS & +0.333 & -0.012 & +4 & +1 & +1 & Different \\
M16 / RF / S & Native & +0.000 & +0.000 & +0 & +0 & +0 & Same \\
M16 / RF / S & JVX APIs/JVS & +0.000 & +0.000 & +0 & +0 & +0 & Different \\
M20 / gt / C & Native & +0.000 & +0.000 & +0 & +0 & +0 & Same \\
M20 / gt / C & JVX APIs/JVS & +0.000 & +0.000 & +0 & +0 & +0 & Same \\
M20 / gt / S & Native & +0.000 & +0.000 & +0 & +0 & +0 & Same \\
M20 / gt / S & JVX APIs/JVS & +0.000 & +0.000 & +0 & +0 & +0 & Same \\
M20 / RF / C & Native & -0.008 & -0.001 & +1 & +0 & +0 & Different \\
M20 / RF / C & JVX APIs/JVS & -0.397 & -0.002 & +2 & +0 & +0 & Different \\
M20 / RF / S & Native & +0.000 & +0.000 & +0 & +0 & +0 & Same \\
M20 / RF / S & JVX APIs/JVS & +0.000 & +0.000 & +0 & +0 & +0 & Same \\
\bottomrule
\end{tabular}

\endgroup
\end{table}

\FloatBarrier
\section{Microscopy inputs and association profiles}\label{app:microscopy}

\begin{table}[!htbp]
\centering
\caption{Microscopy association profiles from one separate instrumented run per configuration. Source frame ranges are inclusive and zero-based. The largest stock-input matrix is selected by area before Pylapy's cost expansion.}
\label{tab:microscopy-profiles}
\begingroup
\small
\setlength{\tabcolsep}{4pt}
\renewcommand{\arraystretch}{1.1}
\begin{tabular}{@{}lrrrrr@{}}
\toprule
 & & & \multicolumn{3}{c}{Profiled association time (ms/frame)} \\
\cmidrule(lr){4-6}
Sequence & Source frame range & Max. stock input matrix & Stock & Capped & Rust \\
\midrule
Fluo-C2DL-MSC & 0--47 & $6\times6$ & 0.058 & 0.053 & 0.042 \\
PhC-C2DH-U373 & 15--114 & $7\times7$ & 0.062 & 0.052 & 0.044 \\
DIC-C2DH-HeLa & 0--83 & $17\times17$ & 0.063 & 0.055 & 0.044 \\
BF-C2DL-MuSC & 1276--1375 & $22\times22$ & 0.067 & 0.056 & 0.045 \\
Fluo-N2DH-GOWT1 & 0--91 & $29\times29$ & 0.072 & 0.064 & 0.047 \\
BF-C2DL-HSC & 1664--1763 & $159\times159$ & 0.354 & 0.159 & 0.084 \\
Fluo-N2DL-HeLa & 0--91 & $366\times364$ & 1.738 & 0.459 & 0.189 \\
PhC-C2DL-PSC & 200--299 & $497\times498$ & 2.588 & 0.691 & 0.241 \\
Hydra & 130--229 & $793\times731$ & 13.921 & 4.884 & 2.473 \\
\bottomrule
\end{tabular}
\endgroup

\end{table}

\paragraph{Inputs and preparation.}
Table~\ref{tab:microscopy-profiles} gives the exact windows for Table~\ref{tab:microscopy}. ByoTrack prepares detections from the Cell Tracking Challenge \texttt{ERR\_SEG} masks. Hydra preparation projects the mean of input channels, normalizes at quantiles 0.01 and 0.999 with smooth clipping at 1.0, and runs CPU wavelet detection with scale 1, $k=2.5$, minimum area 5 and batch size 20. These inputs and preparation settings follow the earlier benchmark~\cite{reme2026cellbenchmark}. The present observations retain source-frame and input hashes; input preparation remains outside all reported timing boundaries.

\paragraph{Association boundary and aggregate.}
The instrumented interval covers the association call, including the Pylapy adapter, solver and conversion of pairs back to tensors. Cost construction and profiling checks are excluded. Each configuration has 816 calls, 807 with nonempty matrices; accumulated association time is divided by all frames in each clip. The main table's aggregate multiplies each unrounded profile mean by its excerpt frame count and sums the nine clip totals. It is one instrumented run per arm/excerpt, not the sum of six whole-linker rounds or an average of excerpt speedups. These profiles are separate from the whole-linker measurements.

\paragraph{Recorded validation.}
Track digests cover identifiers, start frames, parent and merge identifiers, coordinates, missing-position masks and detection indices. For each arm, the six timed runs and its instrumented run agree. Across arms, all 18 comparisons with stock SciPy have matching track and cost digests. Cost digests cover array types, shapes, values and finite limits. For matching streams, the recorded check compares $\sum_{(i,j)\in A}(C_{ij}-L)$ with float64 accumulation; all recorded objective gaps are zero. The saved-data audit checks these records and checksums without regenerating the unavailable costs or tracks.

\FloatBarrier
\section{Protocol sensitivity and indexed supplement}\label{app:sensitivity}

Only this protocol-sensitivity analysis uses the separate full Linux/Python 3.11 session. It contains 374 cases, including 340 single matrices and 192 geometry cases, compared with 302, 268 and 120 in each primary light session. The comparison recomputes summaries from saved rounds; it is not an additional benchmark run or a fifth environment in the primary figures.

Taking the first three of five assignment rounds changes none of the 818 native-versus-SciPy practical classifications. Across all ten three-round subsets, at most two classifications change. Truncating application rounds from nine to the first three gives maximum observed speedup shifts that round to 0.71\% for tracker updates and 0.38\% for replay. These are conditional sensitivity checks on one fixed full session, not equivalence tests. The saved-data supplement also lists the coverage removed by the geometry setting and baseline rules and identifies expensive admitted baselines without extrapolating omitted arms.

The indexed supplement is organized as S1 protocol sensitivity, S2 complete primary conditions and native classifications, S3 policy/layout timing and diagnostic work, S4 applications, S5 batch/workspace and memory, S6 LSAPE, and S7 the separate microscopy session. Records preserve environment, case, method, reference, source file and available attempt/input/build identities. Protocol exclusions retain their reason rather than becoming zero latency or a win. The generator validates coverage and joins before drawing; the four environments remain separate observations throughout.

\FloatBarrier
\section{Classification tables}\label{app:classifications}

The following tables retain the exact descriptive counts behind the distribution plots. They use the same unrounded-ratio classification and keep environments separate. Counts describe case coverage, not effect magnitude or statistical uncertainty.

\begin{table}[!htbp]
\centering
\caption{Native optional wins/ties/losses by workload family and environment. Each environment contains 166 optional cases. Required JVRUST/JVHYBRID/JVCPP each win all 102 required cases. Family denominators are retained; environments are not pooled.}
\label{tab:nativefamilies}
\begingroup\footnotesize
\setlength{\tabcolsep}{3pt}
\renewcommand{\arraystretch}{1.12}
\begin{tabular}{lrrrrrrrr}
\toprule
& \multicolumn{2}{c}{L311} & \multicolumn{2}{c}{L314} & \multicolumn{2}{c}{W311} & \multicolumn{2}{c}{W314} \\
Family ($n$) & Rust & C++ & Rust & C++ & Rust & C++ & Rust & C++ \\
\midrule
Base (5) & 5/0/0 & 5/0/0 & 5/0/0 & 5/0/0 & 5/0/0 & 5/0/0 & 5/0/0 & 5/0/0 \\
Formulations (5) & 5/0/0 & 4/1/0 & 5/0/0 & 4/1/0 & 4/1/0 & 5/0/0 & 4/1/0 & 5/0/0 \\
Scaling (27) & 27/0/0 & 26/1/0 & 27/0/0 & 26/1/0 & 26/1/0 & 27/0/0 & 27/0/0 & 27/0/0 \\
Large square (4) & 4/0/0 & 4/0/0 & 4/0/0 & 4/0/0 & 4/0/0 & 4/0/0 & 4/0/0 & 4/0/0 \\
Large rectangle (6) & 6/0/0 & 6/0/0 & 6/0/0 & 6/0/0 & 6/0/0 & 6/0/0 & 6/0/0 & 6/0/0 \\
Extreme rectangle (6) & 6/0/0 & 6/0/0 & 6/0/0 & 6/0/0 & 6/0/0 & 6/0/0 & 6/0/0 & 6/0/0 \\
Congestion (5) & 4/1/0 & 4/0/1 & 4/1/0 & 4/0/1 & 4/0/1 & 4/1/0 & 4/0/1 & 5/0/0 \\
Forbidden (4) & 4/0/0 & 4/0/0 & 4/0/0 & 4/0/0 & 4/0/0 & 4/0/0 & 4/0/0 & 4/0/0 \\
Ties (4) & 4/0/0 & 4/0/0 & 4/0/0 & 4/0/0 & 4/0/0 & 4/0/0 & 4/0/0 & 4/0/0 \\
Signed (2) & 2/0/0 & 2/0/0 & 2/0/0 & 2/0/0 & 2/0/0 & 2/0/0 & 2/0/0 & 2/0/0 \\
Rejection (22) & 22/0/0 & 22/0/0 & 22/0/0 & 22/0/0 & 22/0/0 & 22/0/0 & 22/0/0 & 22/0/0 \\
Geometry (76) & 76/0/0 & 76/0/0 & 76/0/0 & 76/0/0 & 76/0/0 & 76/0/0 & 76/0/0 & 76/0/0 \\
\bottomrule
\end{tabular}

\endgroup
\end{table}
\FloatBarrier

\begin{table}[!htbp]
\centering
\caption{Every workload with at least one optional native practical tie or loss, with reference/method ratios by environment. All three use congested-block costs but belong to distinct formulation, scaling and congestion families. Classification uses unrounded ratios; displayed values are rounded.}
\label{tab:nativeexceptions}
\begingroup\footnotesize
\setlength{\tabcolsep}{3pt}
\renewcommand{\arraystretch}{1.12}
\begin{tabular}{llrrrr}
\toprule
Workload & Method & L311 & L314 & W311 & W314 \\
\midrule
Formulation blocks, $256\times320$ & JVRUST & 1.070 & 1.051 & 0.971 & 0.977 \\
Formulation blocks, $256\times320$ & JVCPP & 1.017 & 1.005 & 1.107 & 1.096 \\
Scaling blocks, $128\times160$ & JVRUST & 1.073 & 1.101 & 0.994 & 1.030 \\
Scaling blocks, $128\times160$ & JVCPP & 1.016 & 1.013 & 1.103 & 1.163 \\
Congestion 041, $256\times320$ & JVRUST & 0.984 & 0.985 & 0.916 & 0.926 \\
Congestion 041, $256\times320$ & JVCPP & 0.956 & 0.943 & 1.023 & 1.036 \\
\bottomrule
\end{tabular}

\endgroup
\end{table}
\FloatBarrier

\begin{table}[!htbp]
\centering
\caption{LSAPE wins/ties/losses by family and Windows environment. Counts classify each case's median of three process ratios. $O/P$ is the frozen primary comparison; $O/U$, $U/P$, $P/T$ and $S/T$ retain their distinct controlled interpretations.}
\label{tab:lsapefamilies}
\begingroup\footnotesize
\setlength{\tabcolsep}{3pt}
\renewcommand{\arraystretch}{1.12}
\begin{tabular}{llrrrrrr}
\toprule
Family & Env. & $n$ & O/P & O/U & U/P & P/T & S/T \\
\midrule
Square & W311 & 8 & 0/1/7 & 0/0/8 & 8/0/0 & 0/4/4 & 0/7/1 \\
Square & W314 & 8 & 0/1/7 & 0/0/8 & 8/0/0 & 0/5/3 & 0/8/0 \\
\addlinespace
5:4 uniform & W311 & 16 & 4/2/10 & 4/0/12 & 16/0/0 & 0/9/7 & 0/16/0 \\
5:4 uniform & W314 & 16 & 4/3/9 & 4/0/12 & 16/0/0 & 0/9/7 & 0/16/0 \\
\addlinespace
5:4 congested & W311 & 8 & 8/0/0 & 8/0/0 & 8/0/0 & 0/5/3 & 0/8/0 \\
5:4 congested & W314 & 8 & 8/0/0 & 8/0/0 & 8/0/0 & 0/5/3 & 0/8/0 \\
\addlinespace
4:1 & W311 & 16 & 13/3/0 & 5/1/10 & 16/0/0 & 0/10/6 & 0/16/0 \\
4:1 & W314 & 16 & 13/3/0 & 5/2/9 & 16/0/0 & 0/8/8 & 0/16/0 \\
\addlinespace
Extreme & W311 & 12 & 10/2/0 & 3/3/6 & 12/0/0 & 0/8/4 & 0/12/0 \\
Extreme & W314 & 12 & 10/2/0 & 3/6/3 & 12/0/0 & 0/6/6 & 0/12/0 \\
\addlinespace
Threshold ties & W311 & 7 & 3/4/0 & 0/1/6 & 7/0/0 & 0/7/0 & 0/7/0 \\
Threshold ties & W314 & 7 & 3/4/0 & 0/1/6 & 7/0/0 & 0/5/2 & 0/7/0 \\
\addlinespace
\bottomrule
\end{tabular}

\endgroup
\end{table}
\FloatBarrier

\end{document}